\documentclass[10pt,twocolumn,final]{IEEEtran}  %% SINGLESPACE

\usepackage{amsmath}
\usepackage{amssymb,amsfonts}
\usepackage{epsfig}
\usepackage{psfrag}
\usepackage{color}

\usepackage{enumerate}

\newtheorem{theos}{Theorem}
\newtheorem{props}{Proposition}
\newtheorem{lems}{Lemma}
\newtheorem{cors}{Corollary}

\newtheorem{defns}{Definition}

\newcommand{\blems}{\begin{lems}}
\newcommand{\elems}{\end{lems}}

\newcommand{\btheos}{\begin{theos}}
\newcommand{\etheos}{\end{theos}}

\newcommand{\bprops}{\begin{props}}
\newcommand{\eprops}{\end{props}}

\newcommand{\bdes}{\begin{defns}}
\newcommand{\edes}{\end{defns}}
\newcommand{\bcors}{\begin{cors}}
\newcommand{\ecors}{\end{cors}}

\newcommand{\widgraph}[2]{\includegraphics[keepaspectratio,width=#1]{#2}}

\def \poi {\mathcal{P}}

\newcommand{\spro}{\begin{proof}}
\newcommand{\fpro}{\end{proof}}

\newcommand{\hide}[1]{}

\newcommand{\BadEventA}{{\ensuremath {\mathcal E}}}
\newcommand{\BadEventB}{{\ensuremath {\mathcal B}}}

\newcommand{\eps}{\epsilon}

\newcommand{\beq}{\begin{equation}}
\newcommand{\eeq}{\end{equation}}
\newcommand{\bea}{\begin{eqnarray}}
\newcommand{\eea}{\end{eqnarray}}
\newcommand{\Exs}{\ensuremath{\mathbb{E}}}
\newcommand{\Prob}{\ensuremath{\mathbb{P}}}
\newcommand{\mprob}{\ensuremath{\mathbb{P}}}
\newcommand{\Qprob}{\ensuremath{\mathbb{Q}}}
\newcommand{\acheck}{\ensuremath{a}}

\newcommand{\ibit}{\ensuremath{i}} \newcommand{\jbit}{\ensuremath{j}}
 \newcommand{\Graph}{\ensuremath{G}}
\newcommand{\CheckSet}{\ensuremath{C}}
\newcommand{\Vertex}{\ensuremath{V}}
\newcommand{\code}{\ensuremath{\mathcal{C}}} 
\newcommand{\relaxedpoly}{\ensuremath{\mathcal{P}}}

\newcommand{\expandcoeff}{\ensuremath{\mu}}

\newcommand{\del}{\ensuremath{p}}
\newcommand{\delp}{\ensuremath{q}}
\newcommand{\DirtySet}{\ensuremath{F}}
\newcommand{\bk}{\ensuremath{\backslash}}
\newcommand{\defn}{\ensuremath{: \, = }}
\newcommand{\BitRequestNum}{\ensuremath{X}}
\newcommand{\numgamcheck}{\ensuremath{Z}}
\newcommand{\vdeg}{\ensuremath{d_v}}
\newcommand{\cdeg}{\ensuremath{d_c}}
\newcommand{\Neigh}{\ensuremath{N}}
\newcommand{\DirtySetComp}{\ensuremath{\DirtySet^c}}
\newcommand{\bervar}{\ensuremath{Z}}
\newcommand{\BinDist}{\ensuremath{\operatorname{Bin}}}
\newcommand{\binprob}{\ensuremath{b}}
\newcommand{\numbit}{\ensuremath{n}}
\newcommand{\checknum}{\ensuremath{m}}
\newcommand{\epsone}{\ensuremath{\epsilon_1}} 
\newcommand{\epstwo}{\ensuremath{\epsilon_2}} %

\newcommand{\ReqTail}{\ensuremath{\mathcal{T}}} 

\newcommand{\Rboubar}{\ensuremath{\widebar{y}^{\operatorname{up}}}}
\newcommand{\sbar}{\ensuremath{\widebar{s}}} \long\def\comment#1{}

\newcommand{\sbaronecrit}{\ensuremath{\sbar_{\operatorname{crit}}}}
\newcommand{\gam}{\ensuremath{\gamma_1}}
\newcommand{\gambar}{\ensuremath{\widebar{\gamma}_1}}
\newcommand{\gamtwo}{\ensuremath{\gamma_2}}
\newcommand{\gamtwobar}{\ensuremath{\widebar{\gamma}_2}}
\newcommand{\gamthree}{\ensuremath{\beta}}
\newcommand{\gamthreebar}{\ensuremath{\widebar{\gamthree}}}
\newcommand{\constq}{\ensuremath{\nu}}

\newcommand{\coderate}{\ensuremath{R}}

\newcommand{\Tevent}{\ensuremath{D}}
\newcommand{\neighcrit}{\ensuremath{\bar{\gamma}_{\operatorname{crit}}}}
\newcommand{\Subevent}{\ensuremath{E}}
\newcommand{\Termone}{\ensuremath{U_1}}
\newcommand{\Termtwo}{\ensuremath{U_2}}

\newcommand{\KeyFunc}{\ensuremath{F}} % Key function in the error exponent
\newcommand{\Gfunc}{\ensuremath{G}} % Function to be optimized for the error exponent
\newcommand{\alphacrit}{\ensuremath{\alpha_{\operatorname{crit}}}}
\newcommand{\Vbar}{\ensuremath{\widebar{v}}}
\newcommand{\LI}{\ensuremath{{L_I}}}
\newcommand{\UI}{\ensuremath{{U_I}}}
\newcommand{\gamtwocrit}{\ensuremath{{\gamtwo^{\ast}}}}
\newcommand{\myparagraph}[1]{\subsection{#1}}
\newcommand{\mycode}{\ensuremath{\mathcal{C}}}

\newcommand{\conv}{\ensuremath{\operatorname{conv}}}
\newcommand{\yml}{\ensuremath{\widehat{y}_{\operatorname{ML}}}}
\newcommand{\ylp}{\ensuremath{\widehat{y}_{\operatorname{LP}}}}

\newcommand{\ExpanderEvent}{\ensuremath{\mathcal{G}}}

\newcommand{\Myreq}{\ensuremath{Y}}
\newcommand{\myreq}{\ensuremath{y}}
\newcommand{\myreqvec}{\myreq}

\newcommand{\myreqbar}{\ensuremath{\widebar{\myreq}}}
\newcommand{\myreqopt}{\ensuremath{\widebar{\myreq}}}
\newlength{\widebarargwidth}
\newlength{\widebarargheight}
\newlength{\widebarargdepth}
\DeclareRobustCommand{\widebar}[1]{%
  \settowidth{\widebarargwidth}{\ensuremath{#1}}%
  \settoheight{\widebarargheight}{\ensuremath{#1}}%
  \settodepth{\widebarargdepth}{\ensuremath{#1}}%
  \addtolength{\widebarargwidth}{-0.3\widebarargheight}%
  \addtolength{\widebarargwidth}{-0.3\widebarargdepth}%
  \makebox[0pt][l]{\hspace{0.3\widebarargheight}%
    \hspace{0.3\widebarargdepth}%
    \addtolength{\widebarargheight}{0.3ex}%
    \rule[\widebarargheight]{0.95\widebarargwidth}{0.1ex}}%
  {#1}}

\makeatletter
\def\@cite#1#2{[\if@tempswa #2 \fi #1]}
\makeatother

\makeatletter
\long\def\@makecaption#1#2{
        \vskip 0.8ex
        \setbox\@tempboxa\hbox{\small {\bf #1:} #2}
        \parindent 1.5em  %% How can we use the global value of this???
        \dimen0=\hsize
        \advance\dimen0 by -3em
        \ifdim \wd\@tempboxa >\dimen0
                \hbox to \hsize{
                        \parindent 0em
                        \hfil 
                        \parbox{\dimen0}{\def\baselinestretch{0.96}\small
                                {\bf #1.} #2
                                } 
                        \hfil}
        \else \hbox to \hsize{\hfil \box\@tempboxa \hfil}
        \fi
        }
\makeatother

\begin{document}

\title{Probabilistic Analysis of Linear Programming Decoding}
%
%%%%%%%%%%%%%%%%%%%%%%%%%%%%%%%%%%%%%%%
\author{ ~~~~~~~~~~Constantinos Daskalakis$^1$,\thanks{CD supported by NSF
Grant CCF-0515259; part of this work was done while the author was
visiting Microsoft Research.}
\and Alexandros G. Dimakis$^1$,\thanks{AGD supported by NSF Grants
DMS-0528488, CCF-0635372, and an UC MICRO grant with Marvell
Corporation; part of this work was done while the author was visiting
EPFL.}
\and %\text{~~~~~~~~~~~~~}
\and Richard M. Karp$^1$,\thanks{RMK supported by NSF Grant
CCF-0515259.}
\and and Martin J. Wainwright$^{1,2}$\thanks{MOW supported by an
Alfred P. Sloan Foundation Fellowship, NSF grants DMS-0528488,
CCF-0635372, and an UC MICRO Grant with Marvell Corporation.}}
%
%\thanks{\small $^1$ Department of Electrical Engineering and Computer
%Sciences, UC Berkeley\\$^2$ Department of Statistics, UC Berkeley.}

\maketitle

\begin{abstract}
We initiate the probabilistic analysis of linear programming (LP)
decoding of low-density parity-check (LDPC) codes. Specifically, we
show that for a random LDPC code ensemble, the linear programming
decoder of Feldman et al. succeeds in correcting a constant fraction
of errors with high probability. The fraction of correctable errors
guaranteed by our analysis surpasses previous non-asymptotic results
for LDPC codes, and in particular exceeds the best previous
finite-length result on LP decoding by a factor greater than ten. This
improvement stems in part from our analysis of probabilistic
bit-flipping channels, as opposed to adversarial channels.  At the
core of our analysis is a novel combinatorial characterization of LP
decoding success, based on the notion of a generalized matching.  An
interesting by-product of our analysis is to establish the existence
of ``probabilistic expansion'' in random bipartite graphs, in which one
requires only that almost every (as opposed to every) set of a certain
size expands, for sets much larger than in the classical
worst-case setting.
\end{abstract}

%%%%%%%%%%%%%%%%%%%%%%%%%%%%%%%%%%%%%%%%%%%%%%%%%%%%%%%%%%%%%%%%%%%%%%%%%%
%\newpage

%\setcounter{page}{1}

\section{Introduction} \label{sec:introduction}

%The extraordinary performance of LDPC codes with message passing decoders
%has not been completely understood, since the theoretical tool
%it has been shown that suitably designed LDPC codes, when decoded
%with the sum-product message-passing algorithm~\cite{Frank01}, can
%achieve capacity for erasure channels, and come extremely close
%for more general channels~\cite{Chung01}
% The sparse nature of the graphs defining these codes makes them
%well-suited to iterative message-passing decoding.

Low-density parity-check (LDPC) codes are a class of sparse binary
linear codes, first introduced by Gallager~\cite{Gallager63}, and
subsequently studied extensively by various
researchers~\cite{Mackay99,Richardson01a,Luby01}. 
See the book by Richardson and Urbanke~\cite{mct} for a comprehensive treatment of the subject. 
When decoded
with efficient iterative algorithms (e.g., the sum-product
algorithm~\cite{Frank01}), suitably designed classes of LDPC codes
yield error-correcting performance extremely close to the Shannon
capacity of noisy channels for very large codes~\cite{Chung01}.  Most
extant methods for analyzing the performance of iterative decoding
algorithms for LDPC codes---notably the method of density
evolution~\cite{Luby01,Richardson01a}---are asymptotic in nature,
based on exploiting the high girth of very large random graphs.
Therefore, the thresholds computed using density evolution are only
estimates of the true algorithm behavior, since they assume a
cycle-free message history.  In fact, the predictions of such methods
are well-known to be inaccurate for specific codes of intermediate
block length (e.g., codes with a few hundreds or thousands of
bits). For this reason, our current understanding of practical
decoders for smaller codes, which are required for applications with
delay constraints, is relatively limited.

The focus of this paper is the probabilistic analysis of linear
programming (LP) decoding, a technique first introduced by Feldman
et al.~\cite{Feldman02,Feldman05} as an alternative to iterative
algorithms for decoding LDPC codes.  The underlying idea is a
standard one in combinatorial optimization---namely, to solve a
particular \emph{linear programming} (LP) relaxation of the
integer program corresponding to maximum likelihood (optimal)
decoding.
%
%Feldman et al. proposed a particular LP relaxation of the optimal
%decoding problem, in which the constraint set has a local structure
%defined by the factor graph defining the code (see
%Section~\ref{SecBackground} for further details). More generally,
%there are a hierarchy of LP relaxations associated with the ML
%decoding problem~\cite{Deza97,Barahona86}.
%
Although the practical performance of LP decoding is comparable to
message-passing decoding, a significant advantage is its relative
amenability to non-asymptotic analysis.  Moreover, there turn out to
be a number of important theoretical connections between the LP
decoding and standard forms of iterative
decoding~\cite{KoeVon03,WaiJaaWil05b}. These connections allow
theoretical insight from the LP decoding perspective to be transferred
to iterative decoding algorithms.

\myparagraph{Previous work} The technique of LP decoding was
introduced for turbo-like codes~\cite{Feldman02}, extended to LDPC
codes~\cite{FelKarWai02,Feldman05}, and further studied by various
researchers (e.g.,~\cite{VonKoe06,CheSte06,
Feldman05b,DimWai06,lpcap_soda,Halabi05}).
Significant recent interest has focused on post-processing algorithms that use the ML-certificate property of LP decoding to achieve near ML performance (see \cite{DimWai06, ChertkovEF}) and also
%\cite~{Taghavi07, SDmixed07}.
\cite{SDmixed07, Taghavi07}).

  For concatenated expander
codes, Feldman and Stein~\cite{lpcap_soda} showed that LP decoding can
achieve capacity; see also~\cite{bargzemor2, Kelley07}. For the standard LDPC
codes used in practice, the best positive result from previous
work~\cite{Feldman05b} is the existence of a constant
$\beta > 0$, depending on the rate of the code, such that LP decoding
can correct \emph{any} bit-flipping pattern consisting of at most
$\beta \numbit$ bit flips.  (In short, we say that LP decoding can
correct a $\beta$-fraction of errors.)  As a concrete example, for
suitable classes of rate $1/2$ LDPC codes, Feldman et
al.~\cite{Feldman05b} established that $\beta = 0.000177$ is
achievable.  However, this analysis~\cite{Feldman05b} was worst-case
in nature, essentially assuming an adversarial channel model.  Such
analysis yields overly conservative predictions for the probabilistic
channel models that are of more practical interest.  Consequently, an
important direction---and the goal of this paper---is to develop
methods for finite-length and \emph{average-case analysis} of the LP
decoding method.

\myparagraph{Our contributions} This paper initiates the average-case
analysis of LP decoding for LDPC codes.
%More specifically, our analysis is performed on the
%\emph{bit-regular random ensemble}, in which each one of the $\numbit$
%variable nodes picks a random subset of $\vdeg$ checks.
In particular, we analyze the following question: what is the
probability, given that a random subset of $\alpha \numbit$ bits is
flipped by the channel, that LP decoding succeeds in recovering
correctly the transmitted codeword?  As a concrete example, we prove
that for bit-degree-regular LDPC codes of rate $1/2$ and a random
error pattern with $\alpha \numbit$ bit flips, LP decoding will
recover the correct codeword, with probability converging to
one\footnote{Note that our analysis yields a bound on the probability
of failure for every finite block length $n$.}, for all $\alpha$ up to
at least $0.002$. This guarantee is roughly ten times higher than the
best guarantee from prior work~\cite{Feldman05b}, derived in the
setting of an adversarial channel.  Our proof is based on analyzing
the dual of the decoding linear program and obtaining a simple
graph-theoretic condition for certifying a zero-valued solution of the
dual LP, which (by strong duality) ensures that the LP decoder
recovers the transmitted codeword.  We show that this dual witness has
an intuitive interpretation in terms of the existence of hyperflow
from the flipped to the unflipped bits.  Although this paper focuses
exclusively on the binary symmetric channel (BSC), the poison hyperflow
is an exact characterization of LP decoding for any
memoryless binary input symmetric output (MBIOS) channel. 
 We then
show that such a hyperflow witness exists with high probability under
random errors in the bit-degree-regular LDPC ensemble.  The argument
itself entails a fairly delicate sequence of union bounds and
concentration inequalities, exploiting expansion and matchings on
random bipartite graphs.

\myparagraph{Probabilistic Expanders} 

An interesting by-product of our analysis is
the proof of the existence of \emph{probabilistic expanders}---that is,
bipartite graphs in which \emph{almost all sets} of vertices of
size up to $\alpha \numbit$ \emph{and their subsets} have large expansion. 
One key point is that it is not sufficient to require a random subset of vertices to expand w.h.p., 
because we use the expansion combined with Hall's theorem to guarantee large matchings. What we need instead is that a random subset of vertices and all its subsets will expand w.h.p. 
which by Hall's theorem will guarantee that a random subset will have a matching. 
In effect, by relaxing the expansion requirement from every
set to almost all sets of a given size, we show that one can obtain
much larger expansion factors, and corresponding stronger guarantees
on error correction.
Our analysis relies on the fact that a random bipartite graph, conditioned on all the small sets having some expansion, will also have this probabilistic expansion for much larger constants $\alpha$.  
 This innovation allows us to go beyond the
worst-case fraction of errors guaranteed by traditional expansion
arguments~\cite{Feldman05b,ZipSpi96}.

The remainder of the paper is organized as follows. We begin in
Section~\ref{SecBackground} with background on error-control coding
and low-density parity-check codes, as well as the method of linear
programming (LP) decoding.  Section~\ref{SecMain} describes our main
result and Section~\ref{SecProof} provides the proof in a series of
lemmas, with more technical details deferred to the appendix.  We
conclude in Section~\ref{SecDiscussion} with a discussion.

\section{Background and Problem Formulation}
\label{SecBackground}

We begin with some background on low-density parity-check codes.
We then describe the LP decoding method, and formulate the problem
to be studied in this paper.

\myparagraph{Low-density parity-check codes} The purpose of an
error-correcting code is to introduce redundancy into a data sequence
so as to achieve error-free communication over a noisy channel.  Given
a binary vector of length $k$ (representing information to be
conveyed), the encoder maps it onto a codeword, corresponding to a
binary vector of length $n > k$.  The code rate is given by
$\coderate=k/n$, corresponding to the ratio of information bits to
transmitted bits.  In a binary linear code, the set of all possible
codewords corresponds to a subspace of $\{0,1\}^n$, with a total of
$2^k$ elements (one for each possible information sequence).  The
codeword is then transmitted over a noisy channel.  In this paper, we
focus on the \emph{binary symmetric channel} (BSC), in which each bit
is flipped independently with probability $\alpha$.  Given the
received sequence from the channel, the goal of the decoder is to
correctly reconstruct the transmitted codeword (and hence the
underlying information sequence).

Any binary linear code can be described as the null space of a parity
check matrix $H \in \{0,1\}^{(n-k) \times n}$; more concretely, the
code $\mycode$ is given by the set of all binary strings $x \in
\{0,1\}^n$ such that $H x = 0$ in modulo two arithmetic.  A convenient
graphical representation of such a binary linear code is in terms of
its factor graph~\cite{Frank01}.  The factor graph associated with a
code $\mycode$ is a bipartite graph $\Graph = (\Vertex, \CheckSet)$,
with $\numbit = |\Vertex|$ variable nodes corresponding to the
codeword bits (columns of the matrix $H$), and $\checknum = n - k =
|\CheckSet|$ nodes corresponding to the parity checks (rows of the
matrix $H$). Edges in the factor graph connect each variable node to
the parity checks that constrain it, so that the parity check matrix
$H$ specifies the adjacency matrix of the graph.  A \emph{low-density
parity-check} code is a binary linear code that can be expressed with
a sparse factor graph (i.e. one with $\Theta(1)$ edges per row).

Given a received sequence $\hat{y} \in \{0,1\}^n$ from the BSC, the
maximum likelihood (ML) decoding problem is to determine the closest
codeword (in Hamming distance).  It is well known that ML decoding for
general binary linear codes is NP-hard~\cite{Berlekamp78}, which
motivates the study of sub-optimal but practical algorithms for
decoding.

%\subsection {LP decoding}
\myparagraph{LP decoding} \label{SecLPdecode}

We now describe how the problem of optimal decoding can be
reformulated as a linear program over the \emph{codeword polytope},
i.e. the convex hull of all codewords of the code $\mycode$. For every
bit $\hat{y}_i$ of the received sequence $\hat{y}$, define its
log-likelihood as $ \gamma_i= \log{\left( \frac{ \mprob [ \hat{y_i}
|y_i=0]}{ \mprob [ \hat{y_i} | y_i=1]}\right)}$, where $y_i$
represents the corresponding bit of the transmitted codeword $y$.
%\footnote{For the Binary Symmetric Channel there
%are two possible values for $\gamma_i$.}
Using the memoryless property of the channel,
%the negative
%log-likelihood of a codeword $y$ can be written as
%$
%\sum_{i=1}^{n} (\gamma_i y_i - \log{\Pr[ \hat{y_i} | y_i=0]}).
%$
it can be seen that the maximum likelihood (ML) codeword is
\begin{equation}
\yml \, = \, \arg \min_{y \in \code} \sum _{i=1}^{n} \gamma_i y_i.
\end{equation}
Without changing the outcome of the maximization, we can replace the
code $\code$ by its convex hull $\operatorname{conv}(\code)$, and thus
express ML decoding as the linear program 
\begin{eqnarray}
\label{LP1}
\yml & = & \arg \min_{y \in \text{conv}(\code)} \sum _{i=1}^{n}
\gamma_i y_i.
\end{eqnarray}
Although we have converted the decoding problem from an integer
program to a linear program, it remains intractable because for
general factor graphs with cycles, the codeword polytope does not have
a concise description.

A natural approach, and one that is standard in operations research
and polyhedral combinatorics, is to relax the linear program by taking
only a polynomial set of constraints that approximate the codeword
polytope $\conv(\mycode)$.  The first-order LP decoding
method~\cite{Feldman05} makes use of a relaxation that results from
looking at each parity check, or equivalently at each row of $H$, in
an independent manner. For each check $\acheck \in \CheckSet$ in the
code, denote by $\mycode_\acheck$ the set of binary sequences that
satisfy it---that is, $\mycode_\acheck$ corresponds to the local
parity check subcode defined by check $\acheck$ and its bit neighbors.
Observe that the full code $\mycode$ is simply the intersection of all
the local codes, and the codeword polytope has the \emph{exact}
representation \mbox{$\conv(\mycode) =\conv(
\bigcap_{\acheck=1}^{m}\mycode_\acheck)$.}  The first-order LP decoder
simply ignores interactions between the various local codes, and
performs the optimization over the relaxed polytope given by
\mbox{$\relaxedpoly \defn \bigcap_{\acheck=1}^{m}
\conv(\mycode_\acheck)$.}  Note that $\relaxedpoly$ is a convex set
that contains the codeword polytope $\conv(\mycode)$, but also
includes additional vertices with fractional coordinates (called
\emph{pseudocodewords} in the coding literature).  It can be
shown~\cite{WaiJaaWil05b} that if the LDPC graph had no cycles and
hence were tree-structured, this relaxation would be exact;
consequently, this relaxation can be thought of as tree-based.

In sharp contrast to the codeword polytope for a general factor graph
with cycles, the relaxed polytope $\relaxedpoly$ for LDPC codes is
always defined by a linear number of constraints.  Consequently, LP
decoding based on solving the relaxed linear program
\begin{equation}
\label{EqnRelaxedLP} \ylp = \mathrm{argmin}_{y \in \relaxedpoly}
\sum _{i=1}^{n} \gamma_i y_i,
\end{equation}
can solved exactly in polynomial time using standard LP solvers (e.g.,
interior point or simplex methods), or even faster with iterative
methods tailored to the problem
structure~\cite{FelKarWai02,VonKoe_Turbo,Wainwright02aller,WaiJaaWil05b}.

For completeness, we now provide an explicit inequality description of
the relaxed polytope $\relaxedpoly$. For every check $\acheck$
connected to neighboring variables in the set $N(\acheck)$ and for all
subsets $S \subseteq N(\acheck)$, $|S|$ odd, we introduce the
following constraints
\begin{equation}
\label{inequalities} 
\sum_{i \in N(\acheck)\setminus S} y_i + \sum _{i\in S} (1-y_i) \geq
1.
\end{equation}
Each such inequality corresponds to constraining the $\ell_1$ distance
of the polytope from the sequences not satisfying check
$\acheck$---the forbidden sequences---to be at least one. It can be
shown that these \emph{forbidding inequalities} do not exclude valid
codewords from the relaxed polytope.  We also need to add a set of $2
n$ \emph{box inequalities}---namely, $0 \leq y_i \leq 1$---in order to
ensure that we remain inside the unit hypercube. The set of forbidding
inequalities along with the $[0,1]$-box inequalities define the
relaxed polytope.

Note that, given a check $a$ of degree $\cdeg$, there are
$2^{\cdeg-1}$ local forbidden sequences, i.e. sequences of bits in the
check neighborhood $N(a)$ that do not satisfy the check $a$.
Consequently, for a constant check degree code, the total number of
local forbidden sequences is $2^{\cdeg -1 }m$, so that number of
forbidding inequalities scales linearly in the block length
$n$. Fortunately, in the case of low-density parity-check codes, the
degree $\cdeg$ is usually either a fixed constant (for regular
constructions) or small with high probability (for irregular
constructions) so that the number of local forbidden sequences remains
small. Overall, in the cases of practical interest, the relaxed
polytope can be characterized by a linear number of inequalities in
the way that we have described.  (We refer the interested reader to
\cite{Feldman05, FCasc07} for alternative descriptions more
suitable for the case of large $\cdeg$.)

\section{Main result and proof outline}
\label{SecMain}

In this section, we state our main result characterizing the
performance of LP decoding for a random ensemble of LDPC codes, and
provide an outline of the main steps.  Section~\ref{SecProof}
completes the technical details of the proof.

\myparagraph{Random code ensemble} We consider the random ensemble of
codes constructed according to the following procedure.  Given a code
rate $\coderate \in (0,1)$, form a bipartite factor graph $\Graph =
(\Vertex, \CheckSet)$ with a set of $\numbit = |\Vertex|$ variable
nodes, and $\checknum = |\CheckSet| =
\lfloor(1-\coderate)\numbit\rfloor$ check nodes as follows: (i) Fix a
variable degree $\vdeg \in \mathbb{N}$; and (ii) For each variable
$\jbit \in \Vertex$, choose a random subset $\Neigh(\jbit)$ of size
$\vdeg$ from $\CheckSet$, and connect variable $\jbit$ to each check
in $\Neigh(\jbit)$.  For obvious reasons, we refer to the resulting
ensemble as the \emph{bit-degree-regular} random ensemble, and use
$\code(\vdeg)$ to denote a randomly-chosen LDPC code from this
ensemble.

The analysis of this paper focuses primarily on the binary symmetric
channel (BSC), in which each bit of the transmitted codeword is
flipped independently with some probability $\alpha$.  By
concentration of measure for the binomial distribution, it is
equivalent (at least asymptotically) to assume that a constant
fraction $\alpha \numbit$ of bits are flipped by the channel.  Let
$\Prob$ denote the joint distribution, over both the space of
bit-degree-regular random codes, and the space of $\alpha \numbit$ bit
flips.  With the goal of obtaining upper bounds on the LP error
probability $\Prob[\mbox{LP fails}]$, our analysis is based on the
\emph{expansion} of the factor graph of the code. Specifically, the
factor graph of a code with blocklength $n$ is a $(\expandcoeff,
\del)$-\emph{expander} if all sets $S$ of \emph{variable nodes}, of
size $|S|\leq \expandcoeff n$, are connected to at least $\del |S|$
checks.\footnote{Throughout this paper, we work with codes with simple
parity check constraints (LDPC codes) which are different from the
generalized expander codes \cite{ZipSpi96},\cite{lpcap_soda}
that can have large linear subcodes as constraints.}

\myparagraph{Statement of main result} Our main result is a novel
bound on the probability of error for LP decoding, applicable for
finite block length $n$ and the bit-degree-regular LDPC ensemble.  The
main idea is to show that, under certain expansion properties of the
code, LP decoding will succeed in recovering the correct codeword with
high probability.  We note that a random graph will have the required
expansion properties with high probability.

In particular, we show that for the joint distribution over random
expander bit-degree-regular codes and $\lceil\alpha_{c} \numbit\rceil$
(or less) bit flips by the channel, there exists a constant
$\alpha_{c}$, depending on the expansion properties of the ensemble,
such that LP decoding succeeds with high probability. More formally,
%For a rate $1/2$ code,
%the fraction of correctable errors that we establish, denoted by
%$\alpha_{c}$, is at least ten times higher than the previously known
%result~\cite{Feldman05b} (which holds for adversarial bit-flipping errors). More formally,
%
\btheos
\label{ThmMain}
For every bit-degree-regular LDPC code ensemble with parameters
$\coderate,\vdeg,n$, we specify quadruples $(\alpha_c, c,
\expandcoeff,\del)$ for which the LP decoder succeeds with high
probability over the space of $(\expandcoeff,\del)$-expander
bit-regular random codes and at most $\lceil\alpha_c \numbit\rceil$
bit flips.  The probability of failure decreases exponentially in
$c$---namely
\begin{equation}
\label{eq:hehe}
\Prob\left[~\mbox{\em LP success} \, \mid \, \code(\vdeg) \,
\mbox{\emph{is an $(\expandcoeff,\del)$ expander}} \right] \geq
1-e^{-c\numbit}. \qquad
\end{equation}
\etheos
\noindent 
We note that any factor graph sampled from the bit-regular ensemble
will be an expander with high probability.  In general, the fraction
of correctable errors $\alpha_c$ guaranteed by Theorem~\ref{ThmMain}
is a function of the code ensemble, specified by the code rate
$\coderate$, the bit degree $\vdeg$, its expansion parameters
$\expandcoeff$ and $\del$ and the error exponent $c$.  For any code
rate, the maximum fraction of correctable errors $\alpha_c$ achieved
by our analysis is provably larger than that of the best previously
known result~\cite{Feldman05b} for LP decoding, which guaranteed
correction of a fraction $\frac{ 3 \del -2}{2\del -1} \, \expandcoeff$
of errors.  As a particular illustration of the stated
Theorem~\ref{ThmMain}, we have the following guarantee for rate $R =
\frac{1}{2}$ codes:
\bcors
\label{cor:nonExpanderCase}
For code rate $\coderate=\frac{1}{2}$, bit degree $\vdeg=8$ and error
fraction $\alpha \in (0,0.002)$, the LP decoder succeeds with
probability $1-o(1)$ over the space of bit-degree-regular random codes
of degree $\vdeg$ and $\lceil\alpha \numbit\rceil$ bit flips.
\ecors
More generally, for any code rate $\coderate$, our analysis in
Section~\ref{SecProof} (see discussion following Lemma~\ref{lemma:A3})
specifies conditions for the bit flipping probability $\alpha_c$ and
the expansion parameters $\expandcoeff$ and $\del$ so that the
condition~\eqref{eq:hehe} is satisfied with a suitable choice of error
exponent $c$.

\subsection{Outline of main steps}

We now describe the main steps involved in the proof of
Theorem~\ref{ThmMain}.

\subsubsection{Hyperflow witness}
As in previous work~\cite{Feldman05b}, we prove that the LP decoder
succeeds by constructing a \emph{dual witness}: a dual feasible vector
with zero dual cost, which guarantees that the transmitted codeword is
optimal for the primal linear program.  Using the symmetry of the
relaxed polytope, it can be shown~\cite{Feldman05} that the failure or
success of LP decoding depends only on the subset of bits flipped by
the channel and \emph{not} on the transmitted codeword.  Consequently,
we may assume without loss of generality that the all zero codeword
was transmitted. Moreover, note that, for the binary symmetric channel
(BSC) with flip probability $\epsilon$, the log-likelihood of
each received bit is either
$\log{\left(\frac{1-\epsilon}{\epsilon}\right)}$ or
$-\log{\left(\frac{1-\epsilon}{\epsilon}\right)}$. Since the optimum
of the primal is not affected by rescaling, we may assume without loss
of generality that all $\gamma_i$ are either $1$ or $-1$. Then, every
flipped bit $i$ will be assigned $\gamma_i = -1$, whereas every
unflipped bit $\gamma_i=1$. Under these assumptions, Feldman et
al. \cite{Feldman05b} demonstrated that a dual witness can be
graphically interpreted as a set of weights on the edges of the factor
graph of the code:
\blems[Dual witness~\cite{Feldman05b}]
\label{LemDualWit}
Suppose that all bits in the set $\DirtySet$ are flipped by the
channel, whereas all bit in the complementary set
$\DirtySetComp:=V\setminus \DirtySet$ are left unchanged. Set
$\gamma_i=-1$, for all $i \in \DirtySet$, and $\gamma_i=1$, for all
$i\in \DirtySetComp$. Then linear programming (LP) decoding succeeds
for this error pattern if there exist weights $\tau_{\ibit\acheck}$
for all checks $\acheck \in \CheckSet$ and distinct adjacent bits $\ibit \in
N(\acheck)$ such that the following conditions hold:
\begin{subequations}
\label{EqnDualWit}
\begin{eqnarray}
\label{EqnDualWitA}~~~~~~~\tau_{\ibit\acheck} +
\tau_{\jbit\acheck} & \geq & 0 \qquad \mbox{$\forall$ checks $\acheck
\in \CheckSet$, and}  \\
& & \mbox{$\forall $ adjacent bits $\ibit,\jbit \in N(\acheck)$.}
\nonumber \\
\label{EqnDualWitB} \sum_{\acheck \in N(\ibit)}
\tau_{\ibit\acheck} & < & \gamma_\ibit \qquad \mbox{$\forall \; i \in
V$}.
\end{eqnarray}
\end{subequations}
\label{lemma:feldman et al characterization of LP success} \elems

We next introduce a sufficient condition for the success of LP
decoding, one which is equivalent but arguably more intuitive than the
dual witness definition:
\bdes
\label{DefnHyper}
A {\em hyperflow for $\gamma$} is a set of edge weights $\tau_{ij}$
that satisfy condition~\eqref{EqnDualWitB} and moreover, have the
following additional property: for every check $j \in C$, there exists
a $\poi _j \geq 0$ such that for exactly one variable $i \in N(j)$,
$\tau_{ij}= -\poi_j$ and for all the other $i' \in N(j) \setminus
\{i\}$, $\tau_{i'j}=\poi_j$.
\edes

\begin{figure}[h]
\begin{center}
\widgraph{0.45\textwidth}{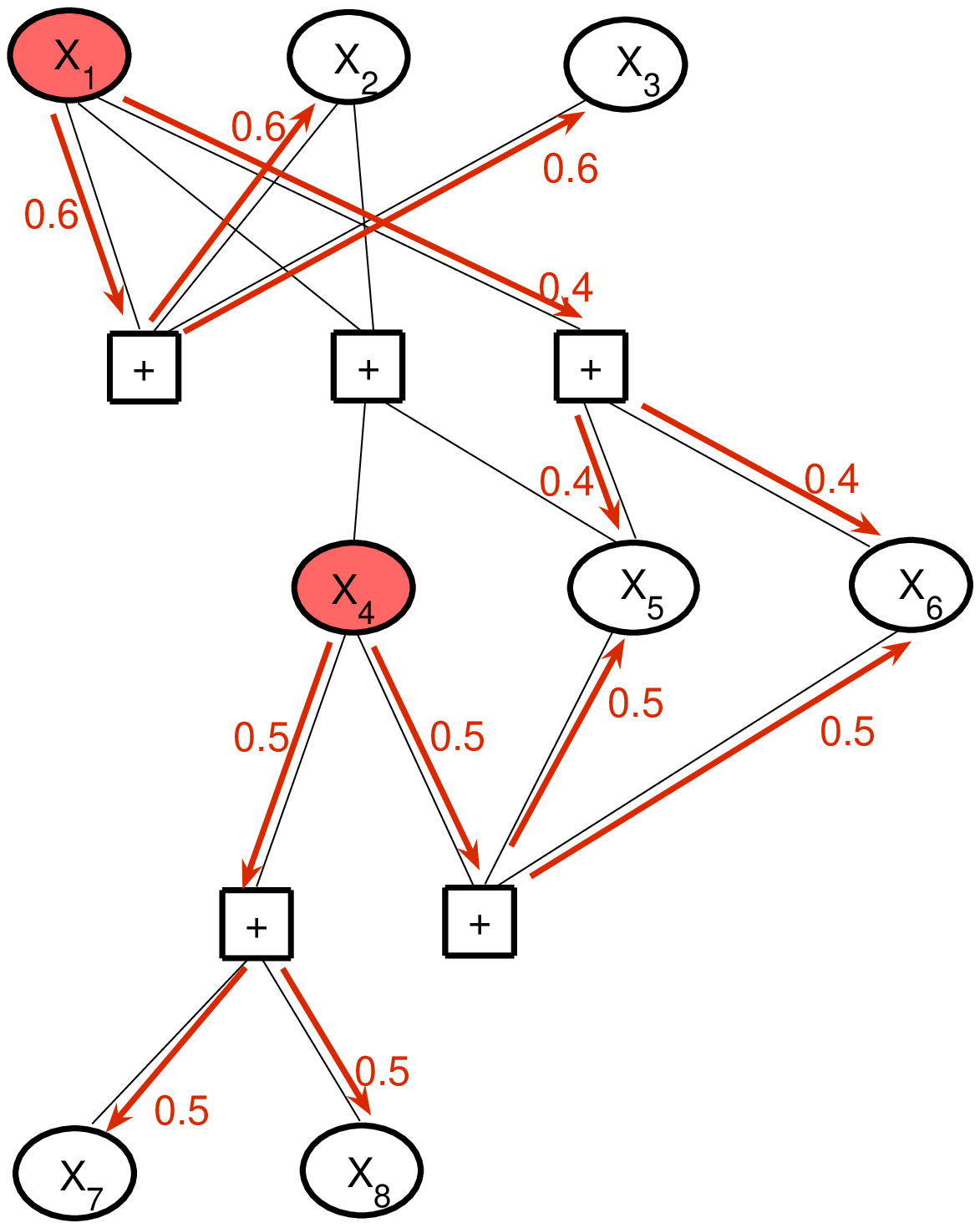}
\vspace{-1cm}
 \caption{Example of a valid hyperflow: bits $x_1$ and $x_4$ have been
 flipped in a binary symmetric channel, and hence each are
 contaminated with one unit of poison.  Each of the unflipped bits can
 absorb up to one unit of poison, whereas the checks act as
 hyper-edges and replicate any incoming flow in all directions other
 than the incoming one. The valid hyperflow shown in this figure
 certifies that LP decoding can correct these two flipped bits. }
 \label{Hyperflow}
\end{center}
\end{figure}

 The flow interpretation is that each check corresponds to a hyperedge
connecting its adjacent variables; the function of any check is to
replicate the flow incoming from one variable towards all its other
adjacent variables.  With this set-up, condition~\eqref{EqnDualWitB}
corresponds to the requirement that all the variables $i$ with
$\gamma_i < 0$ need to get rid of at least $-\gamma_i$ units of
``poison'', whereas each variable $i$ with $\gamma_i>0$ can absorb at
most $\gamma_i$ units of ``poison''.  Figure~\ref{Hyperflow}
illustrates a valid hyperflow for a simple code.

We claim that the existence of a valid hyperflow is equivalent to
the dual witness:
\bprops
\label{PropHyperflow}
There exists a weight assignment $\tau_{ij}$ satisfying the conditions
of Lemma \ref{lemma:feldman et al characterization of LP success} if
and only if there exists a hyperflow $\tau'_{ij}$ for $\gamma$.
\eprops
\noindent See Appendix~\ref{AppPropHyperflow} for a proof of this
claim.

\subsubsection{Hyperflow from $(\del, \delp)$ matching}

\begin{figure}[t]
\begin{center}
\psfrag{#S#}{$\DirtySet$} \psfrag{#nS#}{$N(\DirtySet)$}
\psfrag{#Scomp#}{$\DirtySetComp$}
\widgraph{.42\textwidth}{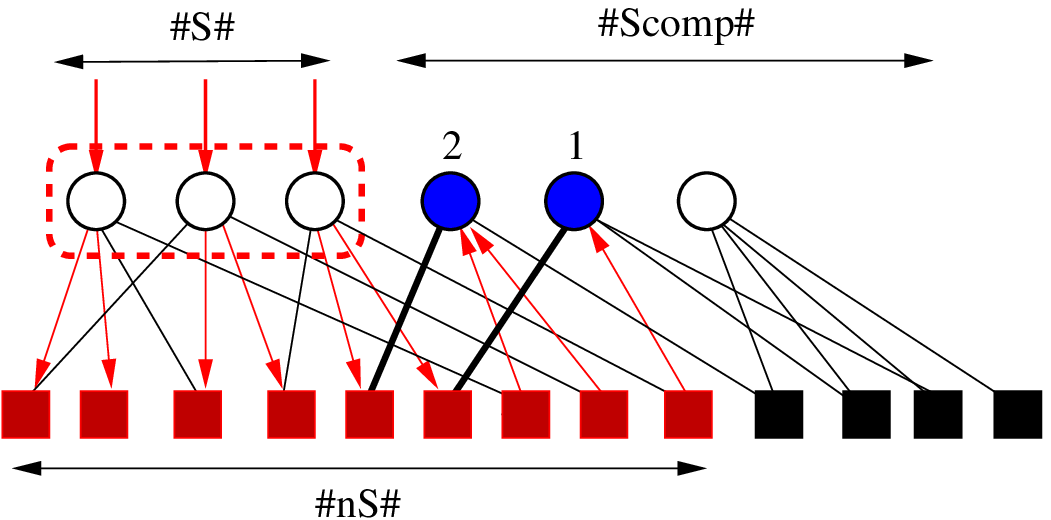}
\caption{Illustration of a $(\del, \delp)$-generalized matching with
\mbox{$\del = 2$,} $\delp = 3$ and $\vdeg = 4$.  The first three bits
are flipped, and form the poisoned set $S$; each flipped bit must be
matched with $\del = 2$ checks from its neighborhood (edges drawn with
arrows). The bit node labeled $2$ lies in $S^c$: it connects to $Z_2 =
|N(2) \cap N(\DirtySet)| = 3$ checks within the set $N(\DirtySet)$,
and so must be matched with $\BitRequestNum_2 = \delp - (\vdeg - 3) = 2$
checks from $N(\DirtySet)$ (two incoming arrows).  By construction,
bit $2$ then has $2+1 = \delp$ checks that are not contaminated.
Similarly, bit $1$ connects $Z_1 = 2$ checks from $N(\DirtySet)$, and
so must be matched with $\BitRequestNum_1 = \delp - (\vdeg-2) = 1$ check from
$N(\DirtySet)$ (incoming arrow).  It has a total of $1 + 2 = \delp$
checks that are not contaminated.}
\label{FigGenMatch}
\end{center}
\end{figure}

Let $N(\DirtySet)$ denote the subset of checks that are adjacent to
the set $\DirtySet$ of flipped bits.  One way to construct a hyperflow
for $\gamma$ is to match each bit $i$ in the set $F$ of flipped bits,
with some number of checks, say $\del \le \vdeg$ checks, to which it
has the exclusive privilege to push flow, suppose in a uniform
fashion. This follows because in a matching each check is used at most once. Let us refer to the checks that are actually used in such a
matching as \emph{dirty}, and to all the checks in $\Neigh(\DirtySet)$
as \emph{potentially dirty}. The challenge is that there might be
unflipped variables that are adjacent to a large number of dirty checks, and
hence fail to satisfy condition~\eqref{EqnDualWitB}; i.e. they receive
more flow than they can absorb. Thus, the goal is to construct the
matching of the flipped bits in a careful way so that no unflipped bit
has too many dirty neighbors. The $\delta$-matching witness, used by
Feldman et al.~\cite{Feldman05b}, avoids this difficulty by matching
\emph{all} of the bits adjacent to potentially dirty checks with
$\delta = \del$ checks each. Our approach circumvents this difficulty
using a more refined combinatorial object that we call a $(\del,
\delp)$-matching. For each bit $\jbit \in \DirtySetComp$, let
\mbox{$\numgamcheck_\jbit \defn |\Neigh(\jbit) \cap
\Neigh(\DirtySet)|$} be the number of its edges adjacent to checks in
$\Neigh(\DirtySet)$.
\bdes
\label{def:pqmatching} Given non-negative integers $\del$ and
$\delp$, a $(\del, \delp)-$\emph{matching} is defined by the
following conditions:
\begin{enumerate}[(a)]
\item each bit $\ibit \in \DirtySet$ must be matched with $\del$
distinct checks, and
\item each bit $\jbit \in \DirtySetComp$ must be matched with
\begin{eqnarray}
\BitRequestNum_\jbit & \defn & \max\{\delp - \vdeg + \numgamcheck_\jbit, 0
\}
\end{eqnarray}
distinct checks from the set $\Neigh(\DirtySet)$.
\end{enumerate}
\edes
In all theoretical analysis in this paper, it is technically
convenient to consider only pairs $(\del, \delp)$ such that
\begin{equation*}
\del \, \geq \, \delp, \quad 2\del + \delp > 2 \vdeg, \quad \mbox{and}
 \quad \vdeg \geq \del + 2.
\end{equation*}
(The lone exception is Figure~\ref{FigGenMatch}, which is shown only
for illustrative purposes.)

We refer to the number of checks with which each variable node needs
to be matched as its \emph{request number}. In this language, all
flipped bits have $\del$ requests while each unflipped bit $\jbit$ has
a variable number of requests $\BitRequestNum_\jbit$ which depends on how
many of its edges land on checks which have flipped neighbors.
The following lemma justifies why requests are selected in this way and illustrates 
the key property of the $(\del,
\delp)$-matching:
\blems
\label{pqmatch} A $(\del,\delp)$-matching guarantees that
all the flipped bits are matched with $\del$ checks, and all
the non-flipped bits have $\delp$ or more non-dirty check
neighbors.
\elems
\noindent This fact follows by observing that any unflipped bit
$\jbit$ with $\numgamcheck_\jbit$ edges in $\Neigh(\DirtySet)$ has
$\vdeg-\numgamcheck_\jbit$ clean neighboring checks, and requests
$\delp -(\vdeg-\numgamcheck_\jbit)$ extra checks from the potentially
dirty ones.

Figure~\ref{FigGenMatch} illustrates a generalized matching for a
degree $\vdeg = 4$ factor graph, and $(\del, \delp) = (2,3)$.  Note
that the bit node labeled $2$ has $Z_2 = 3$ neighbors in the
potentially dirty set $N(\DirtySet)$, and so it makes $\BitRequestNum_2 = 3
- (4-3) = 2$ requests for matching.  This ensures that it is connected
to $2+1 = \delp$ checks that are not dirty.  A similar argument
applies to bit $1$, with $Z_1 = 2$ and $\BitRequestNum_1 = 1$.

We next claim that a $(\del, \delp)$-matching is a certificate of LP
decoding success:
\blems
\label{LemStrongerWitness} For any integers $\del$ and $\delp$ such that $2
\del + \delp > 2 \vdeg$, a $(\del, \delp)$-matching can be used to
generate a set of weights $\tau_{\ibit\acheck}$ which constitute a
hyperflow for $\gamma$ and, hence, satisfy the dual witness
conditions~\eqref{EqnDualWit}.
\elems
\spro Each flipped bit is matched with $\del$ checks: suppose it sends
$\chi$ units of poison to each of these checks.  In the worst case,
the remaining $\vdeg-\del$ edges are connected to checks to which
other flipped bits are sending poison.  Therefore, each flipped bit
(in the worst case) can purge itself of $\del \chi -(\vdeg-\del) \chi$
units of its own poison, so that we require that \mbox{$\del \chi
-(\vdeg-\del) \chi > 1$.}

By Lemma~\ref{pqmatch}, each unflipped bits has at least $\delp$
checks that do not send any poison.  In the worst case, then, an
unflipped bit can receive $(\vdeg-\delp) \chi$ units of poison, which
we require to be less than $1$.  Overall, a valid routing parameter
$\chi$ will exist if $\frac{1}{2\del -\vdeg} < \frac{1}{\vdeg-\delp}$,
or equivalently, if $\del + \delp > 2 \vdeg$ as claimed.
\fpro

In fact, it can verified that our combinatorial witness for LP
decoding success is easier to satisfy than the condition used by
Feldman et al.~\cite{Feldman05b}. Our use of this improved witness,
along with our focus on the probabilistic setting, are the two
ingredients that allow us to establish a much larger fraction
$\alpha_c$ of correctable errors.

\subsubsection{From expansion to matching via Hall's theorem}

The remainder (and bulk) of the analysis involves establishing that,
with high probability over the selection of random expander
bit-degree-regular codes and random subsets of $\lceil\alpha
\numbit\rceil$ flipped bits, a $(\del, \delp)$-matching exists, for
suitable values $\delp \le \del$ to be specified later.  It is well
known~\cite{BM01,ZipSpi96} that random regular bipartite graphs will
have good expansion, with high probability:
\blems[Good expansion]
\label{lemma:expander}
For any fixed code rate $\coderate \in (0,1)$, degree $\vdeg$ and
$\del \le \vdeg-2$, there exist constants $\expandcoeff, c>0$ so that
a code $\code(\vdeg)$ from the bit-degree-regular ensemble of degree
$\vdeg$ is a $(\expandcoeff \numbit, \del)$ expander with probability
at least $1-O(1/n)$.  \elems

Therefore, conditioned that the event that the random graph is an
expander, the next step is to analyze the existence of a $(\del,
\delp)$- matching.  We use Hall's theorem~\cite{vanLint}, which in our
context, states that a matching exists if and only if \emph{every
subset of the variable nodes} have (jointly) enough neighbors in
$\Neigh(\DirtySet)$ to cover the sum of their requests.

Given our random graph and channel models, an equivalent description
of the neighborhood choices for each variable $\jbit \in
\DirtySetComp$ is as follows.  Each node $\jbit \in \DirtySetComp$
picks a random number $\bervar_\jbit \in \{0,1,\ldots,\vdeg\}$
according to the binomial distribution
$\BinDist(\vdeg,\frac{|\Neigh(\DirtySet)|}{m})$, and picks a subset of
$\Neigh(\DirtySet)$ of size $\bervar_\jbit$. This subset corresponds
to the intersection of its check neighborhood $\Neigh(j)$ with the
check neighborhood $\Neigh(\DirtySet)$ of the flipped bits.  The
remaining $\vdeg - \bervar_j$ edges from bit $\jbit$ connect to checks
outside $\Neigh(\DirtySet)$.  With this set-up, we now define the a
``bad event'' $\BadEventA$, defined by the existence of a pair
$(S_1,S_2) \in 2^{\DirtySet}\times 2^{\DirtySetComp}$ of sets that
\emph{contracts}, meaning that it has more requests than neighbors, so
that
\begin{multline}
\label{EqnContract}
\left|\Neigh(S_1) \cup \left[\Neigh(S_2) \cap \Neigh(\DirtySet)
\right] \right| < \\ \del |S_1| + \sum_{\jbit \in S_2}
\max\{0,\delp-(\vdeg - \numgamcheck_\jbit) \} \Big\}.
\end{multline}

\begin{figure}[t]
\begin{center}
\widgraph{0.45\textwidth}{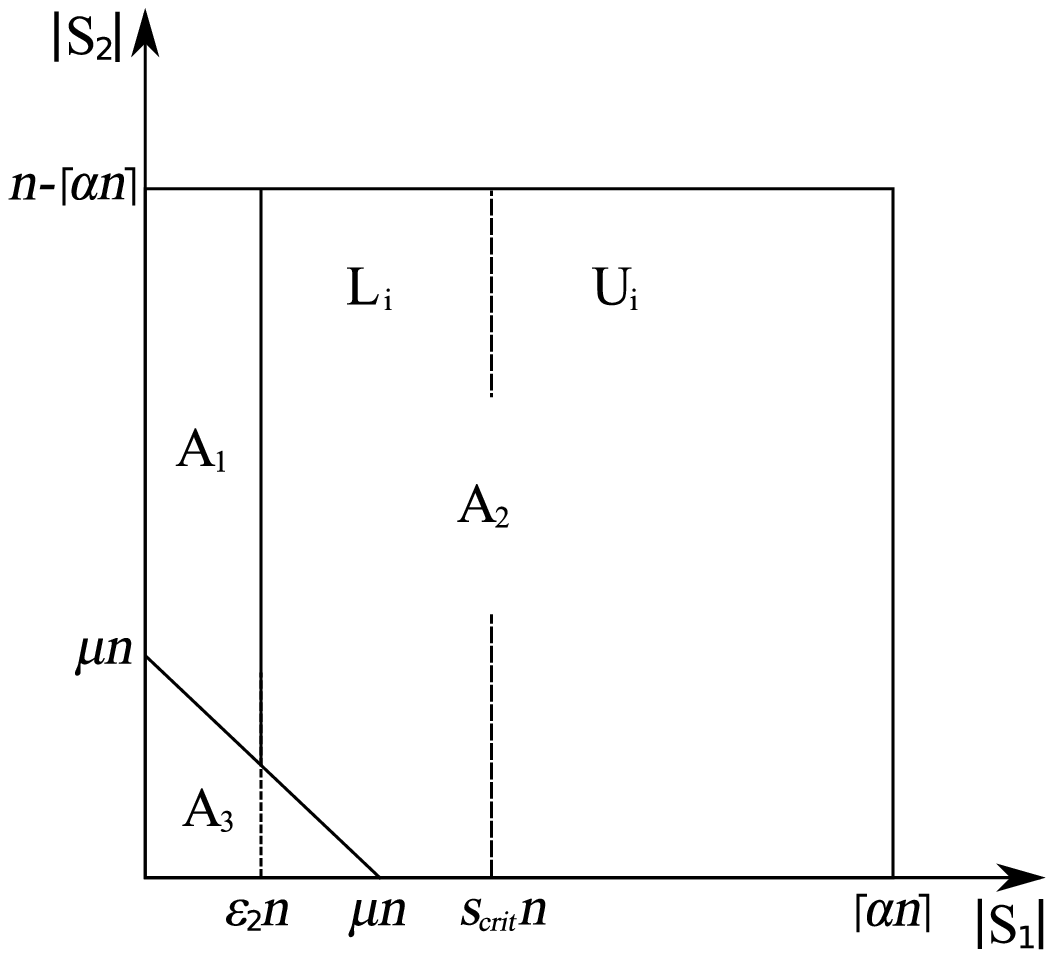}
%\vspace{1cm}
%
\caption{Partitioning the space $2^\DirtySet \times
2^{\DirtySetComp}$.}
\label{Fig:partitionOfSpace}
\end{center}
\end{figure}

Notice that only the neighbors in $\Neigh(\DirtySet)$ are counted,
since a $(\del, \delp)$-matching involves only checks in
$\Neigh(\DirtySet)$. By Lemma~\ref{LemStrongerWitness}, the event
$\BadEventA$ must occur whenever LP decoding fails so that we have the
inequality $\Prob[\mbox{LP decoding fails}] \leq \Prob[\BadEventA]$.
Defining the event 
\begin{eqnarray}
\ExpanderEvent(\vdeg, \expandcoeff, \del) & \defn & \left \{
  \code(\vdeg) \; \mbox{is a $(\expandcoeff n,\del)$ expander} \right
  \},
\end{eqnarray}
we make use of the following conditional form of this inequality:
\begin{eqnarray*}
\Prob\left[~\mbox{LP decoding fails} \, \mid \, \ExpanderEvent \right]
  & \leq & \Prob\left[ \BadEventA \; \mid \; \ExpanderEvent \right].
\end{eqnarray*}
It is useful to partition the space $2^\DirtySet \times
2^{\DirtySetComp}$ into three subsets controlled by the parameters
$\eps_2,\expandcoeff >0$. Parameter $\eps_2 > 0$ is a small constant
to be specified later in the proof and $\expandcoeff$ is the expansion
coefficient.
%specified by Equation~\eqref{eq:expansion} for $\delta =
%\frac{\del}{\vdeg}$. 
The three subsets of interest are given by
\begin{subequations}
\begin{eqnarray}
A_1 & \defn & \{(S_1,S_2) \mid (S_1,S_2)\in A, \; |S_1|+|S_2| <
\expandcoeff n\} \qquad \\
A_2 & \defn & \{(S_1,S_2) \mid (S_1,S_2)\in A-A_1, \; |S_1| \ge \eps_2
n\}, \qquad \\
A_3 & \defn & A \backslash \left(A_1 \cup A_2 \right).
\end{eqnarray}
\end{subequations}
This partition, as illustrated in Figure~\ref{Fig:partitionOfSpace},
decomposes $\BadEventA$ into sub-events
\begin{eqnarray*}
\BadEventA(A_i)& \defn & \left \{ \exists (S_1,S_2)\in A_i \; \Biggr |
\; \mbox{equation~\eqref{EqnContract} holds} \right \}
\end{eqnarray*}
for $i=1,2,3$.  Then, via a series of union bounds, we have the
following upper bound on the probability of failure
\begin{eqnarray*}
\Prob[~\mbox{LP fails} \mid \ExpanderEvent] & \leq & \Prob[~\BadEventA
\; \mid \ExpanderEvent] \\
& \leq & \sum_{i=1}^3 \Prob[~\BadEventA(A_i) \, \mid \, \ExpanderEvent].
\end{eqnarray*}
However, all subsets of variable nodes of size at most $\expandcoeff
n$ in a $(\expandcoeff n,\del)$ expander have a $\del$-matching and,
because $\delp \le \del$, it follows that
\begin{eqnarray}
\label{eq:kostassmallsets}
\Prob[~\BadEventA(A_1) \, \mid \, \ExpanderEvent] & = & 0.
\end{eqnarray}
Consequently, we only have to deal with the remaining two terms of the
summation for $i = 2$ and $i=3$.  Before proceeding, an important
side-remark here is that equation~\eqref{eq:kostassmallsets} by itself
implies that the LP decoder can correct a constant fraction of errors;
indeed, it is precisely this observation that was exploited by Feldman
et al.~\cite{Feldman05b}.  However, our ultimate goal in this paper
is to establish higher fractions of correctable errors, so we need to
continue our analysis further.

For $i=2,3$, we have
\begin{eqnarray}
\Prob \left[\BadEventA(A_i) \; \mid \; \ExpanderEvent \right] & = &
\frac{\Prob[~\BadEventA(A_i)\wedge
\ExpanderEvent]}{\Prob[\ExpanderEvent]} \nonumber \\
& \leq & \frac{\Prob[~\BadEventA(A_i)~]}{\Prob[\ExpanderEvent]}
\nonumber\\
& \leq & 2 \Prob[~\BadEventA(A_i)~],
\end{eqnarray}
where the last inequality follows from Lemma~\ref{lemma:expander}.
Overall, putting everything together, we conclude that
\begin{eqnarray}
\label{middle bound}
\Prob\left[~\mbox{LP fails} \: \mid \; \ExpanderEvent \right] & \leq &
2\sum_{i=2}^3{\Prob[~\BadEventA(A_i)]}.
\end{eqnarray}

The remainder of the proof consists of careful analysis of these two
error terms.  It turns out to be convenient to use an alternative
probabilistic model in the analysis.  In particular, observe that
there is an inconvenient asymmetry in the definition of our
generalized matching: the bits of set $\DirtySetComp$ need to be
matched with checks from the neighborhood of the flipped bits
$\DirtySet$, and not from the whole set of checks from which they
select their neighbors. This correlation between $\Neigh(F)$ and the
number of requests from set $\DirtySetComp$ creates severe
complications in the analysis. Indeed, any attempt to use Hall's
condition through union bounds seems to require independence among
different edges; moreover, crude upper-bounds on the number of
requests from set $\DirtySetComp$ seem inadequate to decorrelate the
requests of $\DirtySetComp$ from the size of $\Neigh(F)$.  For this
reason, we use an alternative probabilistic model, as described in
Section~\ref{section:newProbabilisticModel}.

\section{Proof of Theorem~\ref{ThmMain}}
\label{SecProof}

We now turn to the remaining (somewhat more technical) steps involved
in the proof of Theorem~\ref{ThmMain}.

\subsection{Simplifying the probability model}
\label{section:newProbabilisticModel}

In order to decouple the distribution of the requests of
$\DirtySetComp$ from the size of $\Neigh(\DirtySet)$, observe that the
number of requests $\BitRequestNum_j$ from each bit $j$ in $\DirtySetComp$
grows linearly with the number of edges that this bit has in
$\Neigh(\DirtySet)$. 
Notice that the checks are selected with replacement and the degree of a variable 
can be strictly smaller than $d_v$, although this will not be an issue asymptotically. 
This observation combined with a coupling
argument shows that, if $x, x' \in \{0,\ldots,d_v\}^{|\DirtySetComp|}$
are two vectors of requests from the bits in $\DirtySetComp$, where $x
\leq x'$ elementwise, then the probability that a
$(\del,\delp)$-matching exists is larger conditioned on $x$ than on
$x'$.

%
%In the request computation, we will replace the size of the set of
%potentially dirty checks with its maximum size
%($|\Neigh(\DirtySet)|\leq \alpha \vdeg n$). and further assume
%that the unflipped bits select their neighbors \emph{with
%replacement} in $\Neigh(\DirtySet)$ \footnote{ \textbf{prosoxi!}
%Note that in the selection we use the actual $\Neigh(\DirtySet)$
%set and not its bound, since increasing that would actually
%decrease the probability of contraction.} Clearly, both increasing
%the requests and edge replacement can only increase the
%probability of failure but the benefit is that we can now decouple
%the request selection from the random selections of $\DirtySet$.

This observation suggests the following alternative
experiment:
\begin{itemize}
\item A node $\jbit \in \DirtySetComp$ first picks a random number
\mbox{$\bervar_\jbit \in \{0,1,\ldots,\vdeg\}$} according to the
modified binomial distribution
$\BinDist\left(\vdeg,\frac{\vdeg\lceil\alpha
\numbit\rceil}{m}\right)$.
\item Node $\jbit$ then chooses $\bervar_\jbit$ checks from
$\Neigh(\DirtySet)$ with replacement.
\end{itemize}
This procedure is repeated independently for each $\jbit \in
\DirtySetComp$.  Since $|\Neigh(\DirtySet)| \le \vdeg\lceil\alpha
\numbit\rceil$, the bits of set $\DirtySetComp$ will tend to have more
edges in $\Neigh(\DirtySet)$ and, therefore, more requests in this new
experiment than in the original one (as suggested by the natural
coupling between the two processes). Moreover, since checks are now
chosen with replacement, for each bit $\jbit \in \DirtySetComp$, the
size of the intersection $\Neigh(j) \cap \Neigh(\DirtySet)$ is less
than or equal to $\bervar_\jbit$, since the same check might be chosen
more than once. Intuitively, the existence of matchings is less likely
in the new experiment than in the original one; this claim follows
rigorously by combining these observations with the coupling argument
used in the previous paragraph. The benefit of switching from the
original experiment to this new experiment is in allowing us to
\emph{decouple} the process of deciding the number of requests made by
each bit in $\DirtySetComp$ from the cardinality of the random
variable $\Neigh(\DirtySet)$.

Let us use $\Qprob$ to denote the probability distribution over random
graphs in this new model. Setting $\DirtySetComp(\delp) = \left \{ j
\in \DirtySetComp \; \mid \; \delp > \vdeg - \numgamcheck_\jbit \right
\}$, we can define the alternative ``bad event'' $\BadEventB$, meaning
that there exist $S_1 \subseteq \DirtySet$, and $S_2 \subseteq
\DirtySetComp(\delp)$ such that
\begin{equation}
\label{EqnContract2}
\left|\Neigh(S_1) \cup \left[\Neigh(S_2) \cap \Neigh(\DirtySet)
\right] \right| \leq \del |S_1| + \sum_{\jbit \in S_2}
\left[\delp-(\vdeg - \numgamcheck_\jbit) \right]. \qquad
\end{equation}
In addition, we define the corresponding sub-events $\BadEventB(A_i)$
for $i=1,2,3$. As argued above, it must hold that
\begin{equation*}
\Prob[\BadEventA(A_i)] \leq \Qprob[\BadEventB(A_i)], \quad \mbox{for
all} \quad i = 1, 2, 3,
\end{equation*}
and, therefore, as inequality~\eqref{middle bound} suggests, in order
to upper bound the probability of LP decoding failure, it suffices to
obtain upper bounds on the probabilities $\Qprob[\BadEventB(A_i)]$ for
$i=2,3$. 

For future use, we define for fixed subsets $S_1 \subseteq \DirtySet$
and $S_2 \subseteq \DirtySetComp(\delp)$, the event
$\BadEventB(S_1,S_2)$ that equation~\eqref{EqnContract2} holds for
$S_1$ and $S_2$.  We now proceed, in a series of steps, to obtain
suitable upper bounds on the probabilities $\Qprob[\BadEventB(A_i)]$
and, hence, on the probability of LP decoding failure.

\subsection{Conditioning on requests from $\DirtySetComp$} 
\label{section:mainPROOF}

For each $i \in \{1, \ldots, \delp \}$, we define the random variable
\begin{equation}
\Myreq_i \defn \biggr | \{ \jbit \in \DirtySetComp \; \mid \;
\numgamcheck_\jbit = \vdeg - (\delp - i) \} \biggr |,
\end{equation}
corresponding to the number of bits in $\DirtySetComp$ with $\vdeg -
(\delp -i) $ edges that lie inside the ``contaminated'' neighborhood
$\Neigh(\DirtySet)$ and, hence, with $i$ requests each.  Note that
each $\Myreq_i$ is binomial with parameters $\lfloor(1-\alpha)
\numbit\rfloor$ and
\begin{equation}
\binprob_i \defn {\vdeg \choose {\vdeg-\delp +i}}
\left(\frac{\lceil\alpha\numbit\rceil\vdeg}{m} \right)^{\vdeg - \delp
+i} \left(1-\frac{\lceil\alpha\numbit\rceil\vdeg}{m} \right)^{\delp -
i}.
\end{equation}
Since \mbox{$\Exs[\Myreq_i] = \binprob_i \lfloor(1-\alpha) \,
\numbit\rfloor$}, applying Hoeffding's inequality~\cite{Hoeffding63}
yields the sharp concentration
\begin{equation*}
\Qprob\left[|\Myreq_i - \binprob_i\lfloor(1-\alpha) \, \numbit\rfloor|
\geq \epsone \numbit \right] \; \leq \; 2 \exp \left ( -2 \epsone^2
\numbit \right)
\end{equation*}
for any $\epsone > 0$.  Hence, if we define the event
\begin{equation}
\ReqTail(\epsone) \defn \bigcap_{i=1}^{\delp} \left \{ |\Myreq_i -
\binprob_i\lfloor(1-\alpha) \numbit\rfloor| \leq \epsone \numbit
\right \},
\end{equation}
 then a simple union bound yields that
\begin{equation*}
\Qprob[\ReqTail(\epsone)] \geq 1 - 2 \delp \; \exp \left ( -2
\epsone^2 \numbit \right),
\end{equation*}
 so that it suffices to bound the conditional probabilities
$\Qprob[\BadEventB(A_i) \, \mid \, \ReqTail(\epsone)]$, $i=2,3$.  Note
that conditioned on the event $\ReqTail(\epsone)$, we are guaranteed
that
\begin{eqnarray}
\label{EqnDefnRboubar}
\frac{\Myreq_i}{\numbit} & \leq & \binprob_i (1-\alpha) + \epsone \; =
\,: \; \Rboubar_i.
\end{eqnarray}

We now turn to bounding the probability of the bad event $\BadEventB$.
Since, by symmetry, the probability of the event $\BadEventB(S_1,S_2)$
is the same for different sets $S_1$ of the same size, a union bound
gives $\Qprob[\BadEventB(A_2)\, \mid \, \ReqTail(\epsone)] \leq
\sum_{s_1= \lceil\epstwo \numbit\rceil}^{\lceil\alpha \,
\numbit\rceil} \Tevent(s_1)$, where 
\begin{multline*}
\Tevent(s_1) \defn { \lceil\alpha \numbit \rceil \choose s_1} \times \\
\Qprob \Big[ \exists~S_2 \subseteq \DirtySetComp(\delp) \; \mbox{with}
\; (S_1,S_2) \in A_2 \mbox{ s.t.  } \BadEventB(S_1, S_2) \; \Big |
\ReqTail(\epsone)\Big],
\end{multline*}
with $S_1$ is any fixed set of size $s_1$.

Before bounding these terms, we first partition the values of $s_1$
into two sets $\{\lceil\eps_2\numbit\rceil,\ldots,\lceil\sbaronecrit
\numbit\rceil \}$ and $\{\lceil\sbaronecrit \numbit\rceil+1,\ldots,
\lceil\alpha \numbit\rceil \}$ for some value of $\sbaronecrit$ to be
specified formally in Lemma \ref{LemRegOne}.  To give some intuition,
in the conditional space $\ReqTail(\epsone)$, the total number of
matching-requests from the bits of set $\DirtySetComp$ is at most
\begin{equation}
V \defn \numbit \sum_{i=1}^\delp i \Rboubar_i.
\end{equation}
Therefore, if $\Qprob[\BadEventB(A_2)\, \mid \,\ReqTail(\epsone)]$ is
indeed small, we would expect that, if the set $S_1$ is large enough
(say $|S_1| \approx |\DirtySet|$), then with high probability, the
size of its image $\Neigh(S_1)$ should be large enough not only to
cover its own requests but also $V$ additional
requests---viz. $|\Neigh(S_1)| \ge p|S_1| + V$.  If this condition
holds, then there cannot exist any set $S_2$ such that the event
$\BadEventB(S_1,S_2)$ occurs.  We formalize this intuition in the
following result, proved in Appendix~\ref{AppRegOne}:
\blems[Upper Regime]
\label{LemRegOne}
Define \mbox{$\Vbar \defn \sum_{i=1}^\delp i \;
\Rboubar_i$,} and the function 
\begin{equation*}
f(s) \defn \alpha H\left(\frac{s}{\alpha}\right) +
(1-\coderate)H\left(\frac{\del s + \Vbar}{(1-\coderate)}\right) +
\vdeg s\log_2 \left(\frac{\del s +
\Vbar}{(1-\coderate)}\right),
\end{equation*}
where $H(\cdot)$ is the binary entropy function, and set
\begin{eqnarray}
\label{EqnDefnSbaronecrit}
\sbaronecrit \defn \min \left \{ \alpha, \; \inf \left\{ s \in [0,
\alpha] \; \mid \; f(s') < 0, \forall s' \in [s,\alpha] \right\}
\right \}.
\end{eqnarray}
Then for all $s_1 \in \{\lceil\sbaronecrit \numbit\rceil+1,\ldots,
\lceil\alpha n\rceil\}$, the quantity $\Tevent(s_1)$ decays
exponentially fast in $n$.
\elems
\noindent It remains to bound $\Tevent(s_1)$ for $s_1 \in \LI$, where
\begin{equation*}
\LI \defn \{\lceil\epstwo \numbit\rceil,\ldots, \lceil\sbaronecrit
\numbit\rceil\}. 
\end{equation*}
For a randomly chosen set $S_1$, define the event 
\begin{equation}
\mathcal{F}(s_1, \gam) \defn \{|S_1| = s_1, \quad |\Neigh(S_1)| = \gam
\}.
\end{equation}
By conditioning, we have the decomposition $\Tevent(s_1) =
\sum_{\gam=1}^{\vdeg s_1} \Subevent(\gam, s_1)$, where
\begin{multline*}
\Subevent(\gam, s_1) \defn {\lceil\alpha \numbit\rceil \choose s_1 }
\; \Qprob'[\mathcal{F}(s_1, \gam)] \times \\ \Qprob'\left[ \exists \;
S_2 \mbox{ with } (S_1,S_2)\in A_2 \mbox{ s.t. } \BadEventB(S_1, S_2)
\; \Big| \mathcal{F}(s_1, \gam) \right].
\end{multline*}
Here $\Qprob'$ denotes the conditional probability distribution of
$\Qprob$ conditioned on the event $\ReqTail(\epsone)$.

The following lemma allows us to restrict our attention to
linearly-sized check neighborhoods $\Neigh(S_1)$ in analyzing the
individual terms $\Subevent(\gam, s_1)$ of the summation; the proof is
provided in Appendix~\ref{appendix:LinearSizeNeighborhood}.
\blems[$\gambar$ small]
\label{LemLinSizeCheckNei} 
Define the critical point $\neighcrit(\sbar_1)$
\begin{equation}
\label{EqnSup}
\sup \left \{ \gambar \in (0,\vdeg \sbar_1]
\; \mid \; 2 + \vdeg \sbar_1 \log_2
\left(\frac{\gambar}{(1-\coderate)} \right) < 0 \right\}.
\end{equation}
Then, for set sizes $s_1 \geq \lceil\epstwo \numbit\rceil $ and
neighborhood sizes \mbox{$\gam \leq \neighcrit(\epstwo) \numbit$,} the
quantity $\Subevent(\gam, s_1)$ decays exponentially fast in
$\numbit$.
%$\gam \le
%\neighcrit(s_1/n) \numbit$, the quantity $\Subevent(\gam, s_1)$
%decays exponentially fast in $\numbit$.
%
%
\elems 
\noindent Note that the supremum~\eqref{EqnSup} is always finite.
This lemma essentially says that if $s_1$ has linear size, its neighborhood
$\gam$ must also have linear size. 
\vspace*{.2in}

%\noindent Based on Lemma~\ref{LemLinSizeCheckNei}, it is
%enough to bound $\Subevent(\gam, s_1)$ for values of $\gam$ that
%satisfy $\gam \ge \neighcrit(\epstwo)
%\numbit$.\\

 To summarize our progress thus far, we first argued that in order to
bound the probability $\Qprob[\BadEventB(A_2)\, \mid \,
\ReqTail(\epsone)]$, it suffices to bound the quantities
$\Tevent(s_1)$, for $s_1 \in \{\lceil\epstwo
\numbit\rceil,\ldots,\lceil\alpha \numbit\rceil\}$. Next we
partitioned the range of $s_1$ into two sets: the lower set $\LI =
\{\lceil\eps_2 n\rceil,\ldots,\lceil\sbaronecrit n\rceil \}$, and the
upper set $\UI \defn \{\lceil\sbaronecrit \numbit\rceil+1,\ldots,
\lceil\alpha \numbit\rceil \}$.  The upper set has the property that
for all sets $S_1 \subseteq \DirtySet$ of size $|S_1| \in \UI$, with
high probability the neighborhood $\Neigh(S_1)$ is big enough to
accommodate not only the matching requests from set $S_1$, but also
all possible matching-requests from any set $S_2 \subseteq
\DirtySetComp$. Having established this property of large $S_1$ sets,
it remains to focus on small $S_1$. In this regime, the neighborhood
$\Neigh(S_1)$ on its own is no longer sufficient to cover the joint
set of requests from $S_1$ \emph{and} from any possible set $S_2
\subseteq \DirtySetComp$.  Consequently, one has to consider for every
choice $(S_1, S_2) \in A_2$, whether the joint neighborhood
$\Neigh(S_1) \cup (\Neigh(S_2) \cap \Neigh(\DirtySet))$ is large
enough to cover the matching requests from $S_1$ and $S_2$.

At this point, one might imagine that a rough concentration argument
applied to the sizes of $\Neigh(S_1)$ and $\Neigh(S_2)\cap
\Neigh(\DirtySet)\setminus \Neigh(S_1)$ would suffice to complete the
proof.  Unfortunately, any concentration result must be sufficiently
strong to dominate the factor $\lceil\alpha n\rceil \choose s_1$ that
leads the expression $\Tevent(s_1)$.  Consequently, we study the exact
distribution of the size of $\Neigh(S_1)$, and bound the quantities
$\Subevent(\gam, s_1)$ for $s_1 \in \LI$ and $\gam \in
\{1,\ldots,\vdeg s_1\}$. Of course, since $s_1$ is linear in size, the
bulk of the probability mass is concentrated on linear values for
$\gam$. Therefore, by Lemma \ref{LemLinSizeCheckNei}, we need only
bound $\Subevent(\gam, s_1)$ for $s_1 \in \LI$ and $\gam \ge
\neighcrit(\epstwo)\numbit$.  We complete these steps in the following
subsection.

\subsection{Completing the bound}

Let us fix sizes $s_1 \in \LI$ and $\gam \ge \neighcrit(\epstwo)
\numbit$. For a set $S_1$ of size $s_1$ with neighborhood
$\Neigh(S_1)$ of size $\gam$, define its {\em residual neighborhood}
to be the set $\Neigh(\DirtySet) \bk \Neigh(S_1)$ and use
\mbox{$\gamtwo \defn |\Neigh(\DirtySet) \bk \Neigh(S_1)|$} to denote
its size.  Moreover, define the vector of requests\footnote{Recall
that we have conditioned on the event $\ReqTail(\epsone)$, so that the
number of bits in $\DirtySetComp$ with $i$ matching requests is
concentrated, for every $i\in\{1,\ldots,q\}$.}  $\myreqvec \in
\otimes_{i=1}^{q}\{0,..,\lceil\Rboubar_i\numbit\rceil\}$, let us denote
by $\gamthree(s_1,\gam,\myreq)$ the number of checks missing from the
neighborhood of $S_1$ to cover the total number of requests from $S_1$
and a set $S_2$ with configuration of requests
$\myreq$. Also, let $\constq(\myreq)$ be the number of edges from $S_2$ to $\Neigh(\DirtySet)$. More
precisely, the quantities $\gamthree(s_1,\gam,\myreq)$ and
$\constq(\myreq)$ are given by the formulae
\begin{subequations}
\begin{eqnarray}
\label{EqnDefnGamThree}
\gamthree(s_1,\gam,\myreqvec) & \defn & \del s_1 - \gam +
\sum_{i=1}^{\delp}{i \myreq_i} \\
\constq(\myreqvec) & \defn & \sum_{i=1}^{\delp}{(\vdeg-\delp+i) \,
\myreq_i}.
\end{eqnarray}
\end{subequations}
Note that for $s_1 \in \LI$ and $\gam \ge \neighcrit(\epstwo)
\numbit$, the quantity $\gamthree$ also grows linearly in $\numbit$;
as usual, we use $\gamthreebar$ to denote the rescaled quantity
$\gamthree/\numbit$.  Also recall the definition of $\Rboubar_i$ from
equation~\eqref{EqnDefnRboubar}.  

Letting $\myreqopt \defn (\myreqbar_1, \ldots, \myreqbar_\delp)$ be a
vector of request fractions in $[0,1]^\delp$, we define
\begin{eqnarray*}
\Gfunc(\sbar_1, \gambar, \gamtwobar, \myreqvec) & \defn & \sum_{k=1}^4
\min \{ 0, \Gfunc_k(\sbar_1, \gambar, \gamtwobar, \myreqopt) \}
\end{eqnarray*}
where 
\begin{subequations}
\begin{eqnarray*}
\Gfunc_1 & \defn & \alpha H\left(\frac{\sbar_1}{\alpha}\right)
\sum_{i=1}^{\delp} \Rboubar_i
H\left(\frac{\myreqbar_i}{\Rboubar_i}\right)\\
\Gfunc_2 & \defn & (1-\coderate)
H\left(\frac{\gambar}{(1-\coderate)}\right) + \vdeg \sbar_1
\log_2\left(\frac{\gambar}{(1-\coderate)}\right) \\
\Gfunc_3 & \defn & ((1-\coderate)-\gambar)
H\left(\frac{\gamtwobar}{((1-\coderate)-\gambar)} \right) \\
& & \qquad + \vdeg (\alpha-\sbar_1) \log_2 \left(\frac{\gambar +
\gamtwobar }{(1-\coderate)}\right) \\
\Gfunc_4 & \defn & \gamtwobar
H\left(\frac{\min\{\gamtwobar,\gamthreebar(\sbar_1,\gambar,\myreqbar)\}}{\gamtwobar}\right)
\\
& & \qquad
+ \constq(\myreqbar) \log_2\left(\frac{\gambar+\min\{\gamtwobar,
\gamthreebar(\sbar_1,\gambar,\myreqbar)\} }{\gambar+\gamtwobar}\right)
\end{eqnarray*}
\end{subequations}
With these definitions, we have the following result:
\blems[Exponential upper bound]
\label{LemExpUpper} Suppose that the following inequalities hold:
\begin{subequations}
\label{EqnThreeInequal}
\begin{eqnarray}
\sbaronecrit < \frac{\alpha}{2}, &&\\
\alpha \vdeg < \frac{(1-\coderate)-\vdeg \sbaronecrit}{2}, \quad
\mbox{and} && \\
\alpha H\left(\frac{\sbaronecrit}{\alpha} \right) \vdeg
(\alpha-\sbaronecrit) \log_2 \left(\frac{\vdeg
\sbaronecrit}{(1-\coderate)} \right) \; < \; 0. &&
\end{eqnarray}
\end{subequations}
Then for some $c > 0$, we have the exponential upper bound
\begin{eqnarray*}
\Qprob[\BadEventB(A_2)\, \mid \, \ReqTail(\epsone)] & \leq &
 2^{\numbit \KeyFunc(\alpha)} + \exp(-c n)
\end{eqnarray*}
where the function in the exponent is given by
\begin{eqnarray}
\label{EqnDefnKeyFunc}
F(\alpha) & \defn & \sup_{\sbar_1,\gambar,\gamtwobar,\{\myreqbar_i\}}{
\Gfunc(\sbar_1, \gambar, \gamtwobar,
\myreqbar_1,\ldots,\myreqbar_{\delp})},
\end{eqnarray}
with the maximization over 
\begin{eqnarray*}
\sbar_1 & \in & [0,\sbaronecrit] \\ 
\gambar & \in & [0,\vdeg \sbar_1] \\
\gamtwobar & \in &  [0, \; \vdeg\, (\alpha- \sbar_1) ], \quad \mbox{and} \\
\widebar{y}_i & \in & [\Rboubar_i/2,\Rboubar_i].
\end{eqnarray*}
\elems
\noindent See Appendix~\ref{AppExpUpper} for a proof of this
lemma.

\vspace*{.1in}

It remains to upper bound the probability of the bad-event
$\BadEventB(A_3)$ which is equivalent to the existence of a pair of
contracting sets $(S_1,S_2)$, where the size of set $S_1\subseteq
\DirtySet$ is at most $\epstwo n$ and the size of set $S_2 \subseteq
\DirtySetComp$ is at least $(\expandcoeff - \epstwo) n$. Note that we
haven't yet specified the constant $\epstwo$. The following lemma
establishes that there exists a value of $\epstwo$ so that
$\Qprob[\BadEventB(A_3)\, \mid \, \ReqTail(\epsone)]$ is bounded by an
exponentially decreasing function in $n$ provided that the function
$F(\alpha)$ from equation~\eqref{EqnDefnKeyFunc} is negative.  The
proof of this final lemma is provided in Appendix
\ref{appendix:proofForA3}.

\blems
\label{lemma:A3} If $F(\alpha)<0$, then there exists $\epstwo$
so that the probability $\Qprob[\BadEventB(A_3)\, \mid \,
\ReqTail(\epsone)]$ is decreasing exponentially in $n$.
\elems

We may now complete the proof of Theorem~\ref{ThmMain}.  For a given
rate $\coderate$, fix the bit degree $\vdeg$ and the matching
parameters $(\del, \delp)$ such that
\begin{equation}
\label{EqnCondOne}
2\del+\delp>2\vdeg, \quad \del\ge\delp, \quad \mbox{and} \quad
\vdeg-\del \ge 2,
\end{equation}
 and recall the definition~\eqref{EqnDefnSbaronecrit} of
$\sbaronecrit$.  Suppose that the three
inequalities~\eqref{EqnThreeInequal} hold, and that the function
$F$ defined in equation~\eqref{EqnDefnKeyFunc} satisfies
\begin{eqnarray}
\label{EqnCondThree}
F(\alpha) & < & 0
\end{eqnarray}
Then the probability
\begin{equation*}
\Prob[~\mbox{LP decoding fails} \, \mid \, \code(\vdeg)\mbox{ is a
$(\expandcoeff n,\del)$ expander}~]
\end{equation*}
decays exponentially in $n$, where $\Prob$ is the uniform distribution
over the set of bit-degree-regular codes of degree $\vdeg$ and
selections of $\lceil \alpha n\rceil$ bit flips.

In particular, these explicit conditions allow us to investigate
fractions of correctable errors on specific code ensembles.  As a
concrete example, for code rate $\coderate = 1/2$, if we choose
variable degrees $\vdeg = 8$ and generalized matching parameters
$(\del, \delp) = (6,5)$, one can numerically verify that the
conditions~\eqref{EqnCondOne}, \eqref{EqnCondThree}
and~\eqref{EqnThreeInequal} are satisfied for all $\alpha \leq
\alphacrit = 0.002$. Therefore, for that rate, we establish a fraction
of correctable errors which is more than ten times higher than the
previously known worst-case results, as claimed.

\section{Conclusion}
\label{SecDiscussion}

The main contribution of this paper is to perform probabilistic
analysis of linear programming (LP) decoding of low-density
parity-check (LDPC) codes in the finite-length regime. Specifically,
we showed that for a random LDPC code ensemble, the linear programming
decoder of Feldman et al. succeeds (with high probability) in
correcting a constant fraction of errors that surpasses all prior
non-asymptotic results for LDPC codes.  For a rate $0.5$ code, it
exceeds the best previous finite-length result on LP decoding by a
factor greater than ten.  Despite these substantial improvement, it should be noted that our analysis still yields very
conservative results, roughly a factor of 50 lower than the typical
empirical performance of these codes, as well as the associated
asymptotic thresholds.

Perhaps more important than specific numerical improvements over past
results are the technical innovations that underlie our analysis: a
direct treatment of the probabilistic nature of bit-flipping channels
(as opposed to adversarial analysis in previous work), and a novel
combinatorial characterization of LP decoding, based on the
notion of a poison hyperflow witness.  This hyperflow perspective
illustrates that the factor graph defining a good code should have
good flow properties, in the sense that no matter which subset of bits
are flipped, the poison associated with errors can be diffused and
routed to the unflipped bits. 
For more general MBIOS channels, the amount of poison corresponds exactly to the negative log-likelihood that the channel is assigning to each bit, and the same characterization of LP decoding holds.  

 This intuition suggests that the
property of supporting sufficient hyperflow could provide a useful
design principle in the finite-length setting, for example small sets of variables which contract (are jointly adjacent to few checks) will cause pseudocodewords of small pseuodoweight. 

  There are a number of
ways in which specific technical aspects of the current analysis can
likely be sharpened, which await further work.  In addition, it
remains to further explore the consequences of our analysis technique
for other channels and code ensembles, beyond the particular LDPC
ensemble and binary symmetric channel considered here.

\subsection*{Acknowledgements}

The authors would like the thank the anonymous referees for their 
numerous and detailed suggestions that
helped to improve this paper.  This work was
presented in part in January 2007 at the Symposium on Discrete
Algorithms (SODA), New Orleans, LA.

%%%%%%%%%%%%%%%%%%%%%%%%%%%%%%%%%%%%%%%%%%%%%%%%%%%%%%%%%%%%%%%%%%%%%%%

\bibliographystyle{plain}

\bibliography{soda06_first}

%%%%%%%%%%%%%%%%%%%%%%%%%%%%%%%%%%%%%%%%%%%%%%%%%%%%%%%%%%%%%%%%%%%%%%%

\appendix

\subsection{Proof of Proposition~\ref{PropHyperflow}}
\label{AppPropHyperflow}
%
%%
%\begin{proof}
One direction of the claim is immediate: given the weights
$\tau'_{ij}$ of any hyperflow, they must satisfy
condition~\eqref{EqnDualWitB} by definition, and moreover it is easy
to see that condition~\eqref{EqnDualWitA} will be automatically
satisfied.  In the other direction, we transform the edge weights
$\tau_{ij}$ to new weights $\tau'_{ij}$ that satisfy the hyperflow
constraints. For each check $j$ separately, we replace the weights
$\tau_{ij}$ on the adjacent edges with new weights that satisfy the
hyperflow constraints and, at the same time, do not violate any of the
constraints in condition~\eqref{EqnDualWitB}. Consider a check $j$ and
order the weights on the adjacent edges in increasing order. Assuming
that the check has degree $d(j)$, consider the following cases: \\
\textbf{Case I:} $0\leq \tau_{1j} \leq \tau_{2j} \cdots \leq
\tau_{d(j)j}$. In this case, set $\tau'_{ij}=0$ for all $i$. The new
weights are clearly hyperflow weights and, moreover, it is not hard to
verify that none of the conditions~\eqref{EqnDualWitB} are violated by
the transformation.  \\
\textbf{Case II:} $\tau_{1j} \leq 0 \leq \tau_{2j} \cdots \leq
\tau_{d(j)j}$. Set $\poi_j =-\tau_{1j}$, $\tau'_{1j}=-\poi_j$ and
$\tau'_{i'j}=\poi_j, \forall i' \in N(j)\setminus \{1\}$. This is a
hyperflow weight assignment by construction. Observe, also, that none
of the conditions~\eqref{EqnDualWitB} for the variables in $N(j)$ are
violated by this transformation: indeed, for each variable $k \in
N(j)$, we have that $\tau_{kj} \geq - \tau_{1j}$ since the weights
$\tau_{ij}$ satisfy (\ref{EqnDualWitA}); therefore, setting
$\tau'_{kj}=-\tau_{1j}$ only makes the sum of the edges adjacent to
variable $k$ smaller, and the sum was already satisfying
condition~\eqref{EqnDualWitB} before the transformation.

To conclude the claim, notice that at most one edge adjacent to every
check $j$ can have negative weight in assignment $\tau_{ij}$;
otherwise, condition~\eqref{EqnDualWitB} would be violated for that
check.
%
%it cannot be the case that two or more feasible weight assignments
%$\tau_{ij}, \tau_{i'j}$ are negative, since the sum of every
%pair must be nonnegative. Therefore we have for all the possible
%cases presented a way to construct a valid hyperflow.
$\square$

%%%%%%%%%%%%%%%%%%%%%%%%%%%%%%%%%%%%%%%%%%%%%%%%%%%%%%%%%%%%%%%%%%
%

\subsection{Elementary bounds on binomial coefficients}

\label{AppBin}
For each $\beta \in (0,1)$, define the binomial entropy 
\begin{eqnarray*}
H(\beta) & \defn & - \beta \log_2 \beta - (1-\beta) \log_2 (1-\beta),
\end{eqnarray*}
with $H(0) = H(1) = 0$ by continuity.  We make use of standard
asymptotics of binomial coefficients: for all integers $k$ in the
interval $[0,n]$, we have
\begin{eqnarray}
\label{EqnBinAsymp}
\frac{1}{n} \log {n \choose k}  & = & H(\frac{k}{n}) \pm o(1)
\end{eqnarray}
as $n$ tends to infinity (e.g., see Cover and Thomas~\cite{Cover}).

%%%%%%%%%%%%%%%%%%%%%%%%%%%%%%%%%%%%%%%%%%%%%%%%%%%%%%%%%%%%%%%%%%%%%%%%%%%%%%

\subsection{Proof of Lemma~\ref{LemRegOne}}
\label{AppRegOne}
Note that conditioned on the event $\ReqTail(\epsone)$, we are
guaranteed that $\sum_{i=1}^{\delp}{i \Myreq_i} \leq \Vbar \numbit$.
Letting $S^*_1$ be the fixed subset $\{1, \ldots, s_1 \}$, define the
event $\mathcal{E}(s_1) \defn \left \{ |\Neigh(S^*_1)| \leq \del s_1 +
\Vbar \numbit]\right \}$ and the quantity
\begin{equation*}
P(s_1) \defn {\lceil\alpha \numbit\rceil \choose s_1 }
\Qprob\left[\mathcal{E}(s_1) \right].
\end{equation*}
Using the nature of the bit-regular random ensemble, we have
\begin{eqnarray*}
P(s_1) & \leq & {\lceil\alpha \numbit\rceil \choose s_1}
{\lfloor(1-\coderate) \numbit\rfloor \choose \lfloor\del s_1 + \Vbar
\, \numbit\rfloor} \left(\frac{\lfloor\del s_1 + \Vbar
\numbit\rfloor}{\lfloor(1-\coderate)\numbit\rfloor}\right)^{\vdeg
s_1}.
\end{eqnarray*}
Setting $\sbar_1 = \frac{s_1}{\numbit}$ and using standard bounds on
binomial coefficients~\eqref{EqnBinAsymp}, the quantity $\frac{1}{n}
\log P(s_1)$ is upper bounded by
\begin{align*}
\big [\alpha H \left(\frac{\sbar_1}{\alpha}\right) + (1-\coderate)
H\left(\frac{\del \sbar_1 + \Vbar}{1-\coderate}\right) + \\
 \vdeg \sbar_1
\log_2 \frac{(\del \sbar_1 + \Vbar)}{(1-\coderate)} + o(1)\big ].
\end{align*}
Defining the function $f$ and value $\sbaronecrit$ as in the lemma
statement, we are guaranteed that $P(s_1)$ decays exponentially in $n$
for all $s_1 \in \{\lceil\sbaronecrit n\rceil+1, \ldots, \lceil\alpha
n\rceil\}$. To complete the proof of the claim, we claim that
$\Tevent(s_1)$ can be upper bounded by $P(s_1)$.  Indeed, for $s_1 \in
\{\lceil\sbaronecrit n\rceil+1, \ldots, \lceil\alpha n\rceil\}$, we
can either condition on $\mathcal{E}(s_1)$ or its complement to obtain
that $\Tevent(s_1)$ is upper bounded by
\begin{multline*}
{\lceil\alpha \numbit\rceil \choose s_1} \Biggr \{ \Qprob'[\exists
(S^*_1,S_2) \in A_2 \\ \hbox{ s.t. }\BadEventB(S^*_1, S_2) \, \big | \,
\mathcal{E}^c(s_1) ] 
+ \Qprob[\mathcal{E}(s_1)] \Biggr\},
\end{multline*}
which is equal to $P(s_1)$ because, as argued in Section
\ref{section:mainPROOF}, conditioned on the event
$\mathcal{E}^c(s_1)$, there can be no $S_2$ such that the event
$\BadEventB(S^*_1, S_2)$ holds.
%%%%%%%%%%%%%%%%%%%%%%%%%%%%%%%%%%%%%%%%%%%%%%%%%%%%%%%%%%%%%%%%%%%%%%%%%%%%%%
\subsection{Proof of Lemma~\ref{LemLinSizeCheckNei}}
\label{appendix:LinearSizeNeighborhood}

We have the bound 
\begin{eqnarray*}
\Subevent(\gam, s_1) & \leq & {\lceil\alpha \numbit\rceil \choose s_1
} \Qprob\left[~|\Neigh(S_1)| = \gam~|~|S_1|=s_1 \right],
\end{eqnarray*}
 where we have used the fact that the event $\{|\Neigh(S_1)| = \gam
\}$ is independent of $\ReqTail(\epsone)$ under the probability
distribution $\Qprob$. An exact computation yields that $
\frac{1}{\numbit} \log \left \{ \Qprob\left[|\Neigh(S_1)| =
\gam~|~|S_1|=s_1 \right] \right\}$ is upper bounded by
\begin{equation*}
\frac{1}{\numbit} \log \left \{ {\lfloor(1-\coderate) \numbit\rfloor
\choose \gam} \left(\frac{\gam}{\lfloor(1-\coderate) \numbit\rfloor}
\right)^{\vdeg s_1} \right \},
\end{equation*}
which is in turn upper bounded by
\begin{multline*} 
\ell(s_1) \; \defn \; \Biggr \{
(1-\coderate)H\left(\frac{\gam}{\lfloor(1-\coderate)\numbit\rfloor}\right)
+ \\
\vdeg \frac{s_1}{n} \log_2
\left(\frac{\gam}{\lfloor(1-\coderate)\numbit\rfloor} \right) \Biggr
\} + o(1),
\end{multline*}
where we have used standard bounds on binomial
coefficients~\eqref{EqnBinAsymp}.  Overall, we
have
\begin{eqnarray*}
\frac{1}{\numbit} \log \Subevent(\gam, s_1) & \leq & \alpha
H\left(\frac{s_1}{\lceil\alpha\numbit\rceil} \right) + \ell(s_1) \\
& \leq & \left \{2 + \vdeg \frac{s_1}{n} \log_2
\left(\frac{\gam}{\lfloor(1-\coderate)\numbit\rfloor} \right)
\right\},
\end{eqnarray*}
since $\alpha, \coderate \in (0,1)$, and each entropy term remains
bounded within $[0,1]$.

Finally, setting $\sbar_1 = s_1/\numbit$ and $\gambar = \gam/
\numbit$, consider the function
\begin{eqnarray*}
g(\gambar) & \defn & 2 + \vdeg \sbar_1 \log_2
\left(\frac{\gambar}{(1-\coderate)} \right).
\end{eqnarray*}
We have $\lim_{\gambar \rightarrow 0^+} g(\gambar) = -\infty$,
implying that $\Subevent(\gam, s_1)$ decays exponentially fast in
$\numbit$ for all $s_1 \ge \lceil\epstwo \numbit\rceil $ and
neighborhood sizes $\gam \le \neighcrit(\epstwo) \numbit$, where
$\neighcrit(\cdot)$ is defined as in the statement of the lemma.

%%%%%%%%%%%%%%%%%%%%%%%%%%%%%%%%%%%%%%%%%%%%%%%%%%%%%%%%%%%%%%%%%%%%%

\subsection{Proof of Lemma~\ref{LemExpUpper}}
\label{AppExpUpper}

We begin by proving the following lemma, which provides an upper bound
on the quantity $\Subevent(\gam, s_1)$.
\blems[Lower Regime]
\label{LemRegTwo}
If the three conditions~\eqref{EqnThreeInequal} hold, then, for all
$s_1 \in \{\lceil\epstwo \numbit\rceil,\ldots, \lceil\sbaronecrit
\numbit\rceil\}$ and $\gam \ge \neighcrit(\epstwo) \numbit$, there
exists some $\gamtwocrit = \gamtwocrit(\sbaronecrit,\epstwo)>0$ such
that 
\begin{eqnarray*}
\frac{1}{\numbit} \log \Subevent(s_1, \gam) & \leq &
\sum_{k=1}^3 \min\{0, T_k(s_1, \gam) \} + T_4(s_1, \gam) + o(1),
\end{eqnarray*}
where
\begin{eqnarray*}
T_1 & = & \frac{1}{n} \log {(1-\coderate) \numbit \choose \gam} +
\frac{1}{n} \log \left(\frac{\gam}{(1-\coderate)
\numbit}\right)^{\vdeg s_1} \\
T_2 & = & \frac{1}{n} \log {(1-\coderate) \numbit-\gam \choose \gamtwo
} + \frac{1}{n} \log \left( \frac{\gamtwo+ \gam}{(1-\coderate)
\numbit} \right)^{(\alpha \numbit-s_1) \vdeg} \\
T_3 & = & \frac{1}{n} \log {\gamtwo \choose
\min\{\gamthree(s_1,\gam,y),\gamtwo\}} \nonumber \\
& & \qquad + \frac{1}{n} \log
\left(\frac{\gam+\min\{\gamtwo,\gamthree(s_1,\gam,y)\}}{\gam +
\gamtwo}\right)^{\constq(y)} \\
T_4 & = & \max_{\gamtwo \in \Gamma}
\max_{\myreq_i \in \mathcal{Y}_i} \Biggr [ \sum_{i=1}^\delp
\frac{1}{n} \log {\lceil\Rboubar_i \numbit\rceil \choose \myreq_i} \Biggr]
\end{eqnarray*}
where 
\begin{eqnarray*}
\Gamma & \defn &  \{\lceil\gamtwocrit \numbit\rceil,
\lceil\gamtwocrit \numbit\rceil + 1, \ldots, \vdeg\, (\alpha \numbit-
s_1)\} \\ 
\mathcal{Y}_i & \defn &
\left\{\left\lfloor\frac{\Rboubar_i\numbit}{2}\right\rfloor,\ldots,\lceil\Rboubar_i
\numbit \rceil\right\}, \qquad \mbox{for $i=1, \ldots, \delp$.}
\end{eqnarray*}
\elems
\spro

We begin with the decomposition
\begin{equation}
\label{EqnDecompositionTwo} \Subevent(s_1,\gam) =
{\lceil\alpha \numbit\rceil \choose s_1 } \; \sum_{\gamtwo
=1}^{\vdeg\, \lceil\alpha \numbit\rceil- s_1} \Termone(\gam, \gamtwo)
\Termtwo(\gam, \gamtwo)
\end{equation}
where
\begin{subequations}
%\begin{eqnarray*}
%\Termone(\gam, \gamtwo) & \defn  \Qprob'\big [\exists~S_2\hbox{ with
%}(S_1,S_2)\in A_2 \\
%%
% \hbox{ s.t. }  \BadEventB(S_1, S_2) \;  &| \; 
%|\Neigh(S_1)| = \gam, \; |\Neigh(\DirtySet) \bk \Neigh(S_1)| =
%\gamtwo, |S_1|=s_1 \big ] \\
%%
%\Termtwo(\gam, \gamtwo) & \defn \Qprob'\left[|\Neigh(S_1)| =
%\gam, \; |\Neigh(\DirtySet) \bk \Neigh(S_1)| = \gamtwo \;~|~
%|S_1|=s_1 \right],
%\end{eqnarray*}
%%%---lets try again
\begin{align*}
\Termone(\gam, \gamtwo)  \defn  \Qprob'\big [\exists~S_2\hbox{ with
}(S_1,S_2)\in A_2 \quad \\
 \hbox{ s.t. }  \BadEventB(S_1, S_2) \;  \Big |  
|\Neigh(S_1)| = \gam, \\
 |\Neigh(\DirtySet) \bk \Neigh(S_1)| =
\gamtwo, |S_1|=s_1 \big ], \\
\Termtwo(\gam, \gamtwo)  \defn  \Qprob'\big[|\Neigh(S_1)| =
\gam, \; \hspace{2cm}\\
|\Neigh(\DirtySet) \bk \Neigh(S_1)| =  \gamtwo \big |
|S_1|=s_1 \big]
\end{align*}
\end{subequations}
and recall that $\Qprob'$ is the distribution $\Qprob$ conditioned on
the event $\ReqTail(\epsone)$. We now require a lemma that allows us
to restrict appropriately the range of summation over to values of $\gamtwo$
that scale linearly in $\numbit$.
\blems \label{lemma:smallgamma2}
($\gamtwo$ small): The conditions of Lemma \ref{LemRegTwo} imply that there exists some
value $\gamtwocrit=\gamtwocrit(\sbaronecrit)>0$ for which the quantity
\begin{equation}
G(s_1,\gam):={\lceil\alpha \numbit\rceil \choose s_1 } \;
\sum_{\gamtwo =1}^{\lfloor\gamtwocrit \numbit\rfloor} \Termone(\gam,
\gamtwo) \Termtwo(\gam, \gamtwo)
\end{equation}
decays exponentially in $n$ for any $s_1$ , $\gam$ that satisfy
$\lceil\epstwo \numbit\rceil \le s_1 \le \lceil\sbaronecrit
\numbit\rceil$ and $\gam \ge \neighcrit(\epstwo) \numbit$.  
\elems
\spro 
The proof is similar in spirit to the proof of
Lemma~\ref{LemLinSizeCheckNei}. Taking a term in the
summation~\eqref{EqnDecompositionTwo}, we can bound it as follows:
\begin{eqnarray*}
B(s_1,\gam,\gamtwo) & \defn & {\lceil\alpha \numbit\rceil \choose s_1
 } \; \Termone(\gam, \gamtwo) \Termtwo(\gam, \gamtwo)\\ 
& \leq & {\lceil\alpha \numbit\rceil \choose s_1 }\Termtwo(\gam,
 \gamtwo),
\end{eqnarray*}
which is upper bounded by
\begin{equation*}
{ \lceil\alpha \numbit\rceil \choose s_1 }
\;\Qprob'\left[|\Neigh(\DirtySet) \bk \Neigh(S_1)| =
\gamtwo~~|~~|\Neigh(S_1)| = \gam, |S_1|=s_1 \right].
\end{equation*}
Note that $\Qprob'$ term is upper bounded by
\begin{equation}
\label{EqnQprobBouOne}
{\lfloor(1-\coderate) \numbit\rfloor-\gam \choose \gamtwo}\left(
\frac{\gamtwo+\gam}{\lfloor(1-\coderate) \numbit\rfloor}
\right)^{(\lceil\alpha \numbit\rceil-s_1) \vdeg},
\end{equation}
so that $\frac{1}{n} \log_2 B(s_1,\gam,\gamtwo) \leq \sum_{i=1}^3
C_i(s_1, \gam, \gamtwo) + o(1) $, where
\begin{eqnarray*}
C_1 & = &  \alpha H\left(\frac{s_1/\numbit}{\alpha} \right) \\
C_2 & = & 
H\left(\frac{\gamtwo/\numbit}{(1-\coderate)-\gam/\numbit}\right) \\
C_3 & = & \vdeg \left(\alpha-\frac{s_1}{\numbit}\right) \log_2
\left(\frac{\gamtwo/\numbit+\gam/\numbit}{(1-\coderate)} \right)
\end{eqnarray*}
Using conditions~\eqref{EqnThreeInequal}, since $\sbaronecrit <
\frac{\alpha}{2}$, the term $C_1$ is increasing in $s_1/\numbit$.
Moreover, since $\alpha \vdeg < \frac{(1-\coderate)-\vdeg
\sbaronecrit}{2}$, second entropy term is the term $C_2$ increasing in
$\gam$.  Finally, the term $C_3$ increasing in $s_1/n$ and in
$\gam/n$.

%\\&~~~~~\leq\numbit \left \{\alpha
%H\left(\frac{\sbaronecrit}{\alpha} \right) +
%H\left(\frac{\gamtwo/\numbit}{(1-\coderate)-\vdeg \sbaronecrit}\right)
%+ \vdeg (\alpha-\sbaronecrit) \log_2 \left(\frac{\gamtwo/\numbit+\vdeg
%\sbaronecrit}{(1-\coderate)} \right)+o(1) \right\}.
Consequently, $\frac{1}{n} \log_2 B(s_1,\gam,\gamtwo)$ is upper bounded
by the function
\begin{multline*}
b(\gamma) \defn \alpha H\left(\frac{\sbaronecrit}{\alpha} \right) +
H\left(\frac{\gamma}{(1-\coderate)-\vdeg \sbaronecrit}\right) \\ 
+ \vdeg
(\alpha-\sbaronecrit) \log_2 \left(\frac{\gamma+\vdeg
\sbaronecrit}{(1-\coderate)} \right).
\end{multline*}
Note that $\lim_{\gamma\rightarrow 0}{b(\gamma)}<0$ follows from the
third condition in the series~\eqref{EqnThreeInequal}.  The remainder
of the proof is entirely analogous to that of
Lemma~\ref{LemLinSizeCheckNei}.
 \fpro \hfill 
\noindent By Lemma \ref{lemma:smallgamma2}, it suffices to provide
upper bounds for the terms $B(s_1,\gam,\gamtwo)$ for $s_1 \in
\{\lceil\epstwo \numbit\rceil,\ldots,\lceil\sbaronecrit
\numbit\rceil\}$, $\gam \ge \neighcrit(\epstwo) \numbit$ and $\gamtwo
\ge \gamtwocrit \numbit$. Recall the bound~\eqref{EqnQprobBouOne} on
$\Qprob'\left[|\Neigh(\DirtySet) \bk \Neigh(S_1)| =
\gamtwo~~|~~|\Neigh(S_1)| = \gam, |S_1|=s_1 \right]$.  Similarly,
recall from Appendix~\ref{appendix:LinearSizeNeighborhood} that
$\Qprob'\left[|\Neigh(S_1)| = \gam |S_1|=s_1 \right]$ is upper bounded
by
\begin{equation*}
{\lfloor(1-\coderate) \numbit\rfloor \choose \gam}
\left(\frac{\gam}{\lfloor(1-\coderate) \numbit\rfloor} \right)^{\vdeg
s_1}.
\end{equation*}
Recalling the notation $S^*_1$ for the fixed set $\{1, \ldots, s_1\}$,
he only missing piece is an upper bound on 
\begin{multline*}
\Qprob'\Biggr[\exists (S^*_1,S_2)\in A_2 \hbox{ s.t. }\BadEventB(S^*_1,
S_2) \; \Biggr | \; \\
\Neigh(S^*_1)| = \gam, \; |\Neigh(\DirtySet) \bk
\Neigh(S^*_1)| = \gamtwo \Biggr].
\end{multline*}
Recall that $\Qprob'$ is the conditional probability given the event
$\{\ReqTail(\epsone) \}$.  In this space, every set $S_2 \in
\DirtySetComp(\delp)$ corresponds to a request vector $y \in
\prod_{i=1}^{q}\{0,..,\lceil\Rboubar_i \numbit\rceil\}$. Moreover, for
a set $S_2 \in \DirtySetComp(\delp)$ and its corresponding request
vector $y$, the event $\BadEventB(S^*_1, S_2)$ is equivalent to the
following condition being satisfied:
\begin{equation*}
\BadEventB(S^*_1, S_2) \Leftrightarrow |(\Neigh(S_2)\cap
\Neigh(\DirtySet))-\Neigh(S^*_1)|\le \gamthree(s_1,\gam,y).
\end{equation*}
%\gamthree(s_1,\gam,r) \; \defn \; \del s_1 - \gam + \sum_{i=1}^{\delp}{i
%r_i}, \quad \mbox{and} \quad \constq(r) \;
%\defn \; \sum_{i=1}^{\delp}{(\vdeg-\delp+i)r_i}.
Therefore, a union bound over all the possible choices of sets $S_2$
gives the following upper bound for the probability of interest:
\begin{eqnarray*}
\label{EqnInterThree} 
& & \sum_{\myreq_1=0}^{\lceil\Rboubar_1 \numbit\rceil} \ldots
\sum_{\myreq_{\delp}=0}^{\lceil\Rboubar_{\delp} \numbit\rceil}
\underbrace{{\lceil\Rboubar_1 \numbit\rceil \choose \myreq_1} \ldots
{\lceil\Rboubar_{\delp} \numbit\rceil \choose \myreq_{\delp}} \;
\lambda(\myreq_1,\ldots,\myreq_{\delp},\gam,\gamtwo)}\\ 
& & \qquad \qquad \qquad \qquad \Lambda(\myreq_1,
\myreq_2,\ldots,\myreq_{\delp},\gam,\gamtwo),
\end{eqnarray*}
where $\lambda(\myreq_1,\ldots,\myreq_{\delp},\gam,\gamtwo)$ is the probability,
under the distribution $\Qprob'$, of the event
\begin{multline*}
\Biggr \{ |(\Neigh(S_2)\cap \Neigh(\DirtySet))\setminus\Neigh(S^*_1)|
\leq \gamthree(s_1,\gam,r) \; \Biggr | \; \\
 |\Neigh(S_1)|=\gam, \;
|\Neigh(\DirtySet)\setminus\Neigh(S^*_1)|=\gamtwo \Biggr \},
\end{multline*}
where $S_2$ corresponds to request vector $y$.

\noindent In order to complete the proof, we need a final observation.
\blems 
\label{LemRsize} 
For all $i = 1, \ldots, \delp$, if $\{\myreq_j\}_{j \neq
i},\gam,\gamtwo$ are fixed, then the function
$\Lambda(\myreq_1,\myreq_2,\ldots,\myreq_{\delp},\gam, \gamtwo)$ is
increasing in the scalar variable \mbox{$\myreq_i \in \left\{1, 2,
\ldots,\lfloor\frac{\Rboubar_1 \numbit}{2}\rfloor\right\}$.}
\elems
\spro Clearly ${\lceil\Rboubar_{i} \numbit\rceil \choose \myreq_{i}}$
is increasing for $\myreq_i \in
\left\{1,\ldots,\lfloor\frac{\Rboubar_1
\numbit}{2}\rfloor\right\}$. Therefore, it is enough to establish that
the probability $\lambda(\myreq_1,\ldots,\myreq_{\delp},\gam,\gamtwo)$
is increasing for $\myreq_i \in \left\{ 1, \ldots,
\lfloor\frac{\Rboubar_1 \numbit}{2}\rfloor \right \}$. This fact
follows from the same coupling argument used in
Section~\ref{section:newProbabilisticModel}: for a variable $j \in
\DirtySetComp$, the number of requests $\BitRequestNum_j$ and the size of
the intersection $|\Neigh(j) \cap \Neigh(\DirtySet)|$ are positively
correlated.  Therefore, increasing the number of edges can only
increase the probability
$\lambda(\myreq_1,\ldots,\myreq_{\delp},\gam,\gamtwo)$ of the bad
event $\BadEventB(S_1,S_2)$.
\fpro \hfill 

Using Lemma~\ref{LemRsize}, we can now conclude the proof of
Lemma~\ref{LemRegTwo}.  Denote by
$\mathcal{Y}_i:=\left\{\left\lfloor\frac{\Rboubar_1
\numbit}{2}\right\rfloor,\ldots,\lceil\Rboubar_1
\numbit\rceil\right\}$, we have that
$\frac{1}{n} \log \prod_{i=1}^\delp \Big[\sum_{\myreq_i=0}^{\lceil\Rboubar_i
\numbit\rceil} \Big] \lambda(\myreq_1,\ldots,\myreq_{\delp},\gam,\gamtwo)$
is upper bounded by
\begin{equation*}
\max_{\myreq_i \in \mathcal{Y}_i} \Biggr \{ \sum_{i=1}^\delp \frac{1}{n} \log
{\lceil\Rboubar_i \numbit\rceil \choose \myreq_i} + \frac{1}{n} \log
\lambda(\myreq_1,\ldots,\myreq_{\delp},\gam,\gamtwo) \Biggr \}.
\end{equation*}
By union bound, the quantity $\frac{1}{n} \log
\lambda(\myreq_1,\ldots,\myreq_{\delp},\gam,\gamtwo)$ is upper bounded by
\begin{align*}
\frac{1}{n} \log \Big \{{\gamtwo \choose
\min\{\gamthree(s_1,\gam,r),\gamtwo\} } 
\quad \times \\
\left(\frac{\gam+\min\{\gamthree(s_1,\gam,r),\gamtwo\}} {\gam+\gamtwo}
\right)^{\constq(r)} \Big\}.
\end{align*}
Putting everything together yields the 
claim of Lemma~\ref{LemRegTwo}.  

\fpro \hfill 

Based on the preceding analysis, we can now complete our proof of
Lemma~\ref{LemExpUpper}.  Indeed, using Lemmas~\ref{LemRegOne},
~\ref{LemLinSizeCheckNei}, and~\ref{LemRegTwo} we can upper bound
$\frac{1}{n} \log \Qprob[\BadEventB(A_2)\, \mid \,
\ReqTail(\epsone)]]$.  In this upper bound, all the relevant
quantities (i.e. $s_1$, $\gam$, $\gamtwo$, $\myreq_1, \myreq_2, \ldots,
\myreq_{\delp}$) scale linearly with $n$.  Therefore, standard bounds on
binomial coefficients~\eqref{EqnBinAsymp} lead to the claimed form of
$\KeyFunc$.
%
%\fpro \hfill \qed
%

\subsection{Proof of Lemma \ref{lemma:A3}}
\label{appendix:proofForA3} 
The last thing we need to do is bound the probability of the bad event
$\Qprob'[\BadEventB(A_3)]$ ($S_1$ small, $S_2$ large).
As usual, we do a union bound over $S_1$ sets of various sizes contracting.
Define   
\begin{align*}
\Tevent'(s_1)=
{\lceil\alpha \numbit\rceil \choose s_1} \Qprob' \big [
\exists~S_2 \subseteq \DirtySetComp(\delp) \mbox{ with } (S_1,S_2)\in
A_3 \\ \mbox{ s.t. }\BadEventB(S_1, S_2)\big | \, S_1
\text{ is some fixed set of size } s_1.\big ],
\end{align*}
and therefore 
\begin{align*}
\Qprob'[\BadEventB(A_3)] \leq \sum_{s_1 = 1}^{\lfloor\epstwo \numbit\rfloor}
\Tevent'(s_1).
\end{align*}

Intuitively, it should be clear that this is the easiest regime, because 
 handling requests from $S_1$ is always harder compared to requests from $S_2$ (because variables in $S_2$ have fewer requests). We will make $\epstwo$ small enough so that the requests from $S_1$ are completely covered from the neighborhood of $S_2$ (which is always larger than a linear fraction), 
The function we obtain is strictly dominated by $F(\alpha)$ for sufficiently small $\eps_2$, as one would expect, since $F(\alpha)$ is satisfying the requests in a harder regime. We make a formal argument using continuity to establish this fact. 

For $\eps_2$ sufficiently small, we have that, for all $s_1
\in \{1,\ldots,\lfloor\epstwo \numbit\rfloor\}$,
\begin{equation*}
{{\lceil\alpha \numbit\rceil \choose s_1}} \le {{\lceil\alpha
\numbit\rceil \choose \lfloor\epstwo \numbit\rfloor}} \le n
\left(\alpha H\left(\frac{\eps_2}{\alpha}\right) +o(1)\right).
\end{equation*}
The remainder of the analysis exploits the fact that for $\eps_2$
sufficiently small and any set $S_2$ of size at least $\rho n$, if $y$
is the vector of requests from $S_2$, then, with high probability,
\begin{eqnarray*}
|\Neigh(S_2) \cap (\Neigh(\DirtySet) \setminus \Neigh(S_1))| & \geq &
\underbrace{ \sum_{i=1}^{\delp}{i \myreq_i} + \del\eps_2\numbit}. \\
& & \qquad \gamthree'(\eps_2,y)
\end{eqnarray*}
In words, the neighborhood of set $S_2$ inside $N(F)\setminus N(S_1)$
is sufficiently large not only to cover the requests from set $S_2$
but also from $S_1$. 
We are going to bound the probability of failure, by only allowing
$S_2$ to cover all the requests:
\begin{align*}
\Qprob'[ \exists~S_2 \subseteq \DirtySetComp(\delp)\mbox{ with
}(S_1,S_2)\in A_3 \\
 \mbox{ s.t. }\BadEventB(S_1, S_2) \, \big | \,
S_1\text{ some fixed set of size }s_1] \leq \\
\Qprob'[ \exists~S_2 \subseteq \DirtySetComp(\delp)\mbox{ with
}(S_1,S_2)\in A_3 \\
\mbox{ s.t. }
|\Neigh(S_2) \cap(\Neigh(\DirtySet)-\Neigh(S_1))| \leq \\ \gamthree'(\eps_2,y) 
, \big | \,
S_1\text{ some fixed set of size }s_1]. 
\end{align*}

%%%%%%%%
%
By similar analysis as in the proof of Lemma \ref{LemRegTwo}, we
obtain $\frac{1}{n} \log D'(S_1) \leq F'(\alpha,\eps_2) + o(1)$, where
\begin{eqnarray*}
F'(\alpha,\eps_2) & \defn & \sup_{\gamtwobar \in [0, \;
\vdeg\,\alpha]} \; \sup_{\widebar{y}_i \in [\Rboubar_i/2,\Rboubar_i]}
\Gfunc'(\gamtwobar, \widebar{y}_1, \widebar{y}_2, \ldots,
\widebar{y}_{\delp},\eps_2),
\end{eqnarray*}
and the intermediate function 
\begin{equation*}
\Gfunc' = \Gfunc'(\gamtwobar,
\widebar{y}_1,\ldots,\widebar{y}_{\delp},\eps_2) = \sum_{i=1}^2 \min
\left \{0, \Gfunc'_i(\gamtwobar, \widebar{y}) \right \} +
\Gfunc'_3(\gamtwobar, \widebar{y})
\end{equation*}
has terms
\begin{align*}
\Gfunc'_1  =  ((1-\coderate)-\vdeg\eps_2)
H\left(\frac{\gamtwobar}{((1-\coderate)-\vdeg\eps_2} \right) + \hspace{1cm}\\
 \vdeg
(\alpha-\eps_2) \log_2 \left(\frac{\vdeg\eps_2 + \gamtwobar
}{(1-\coderate)}\right), \\
\Gfunc'_2  =  \gamtwobar
H\left(\frac{\min\{\gamtwobar,\gamthreebar'(\eps_2,\myreqbar)\}}{\gamtwobar}\right) \hspace{3.3cm} \\
+ \constq(\myreqbar) \log_2\left(\frac{\vdeg \eps_2+\min\{\gamtwobar,
\gamthreebar'(\eps_2,\myreqbar)\} }{\vdeg \eps_2+\gamtwobar}\right), \\
\Gfunc'_3  =  \alpha H\left(\frac{\eps_2}{\alpha}\right) +
\sum_{i=1}^{\delp} \Rboubar_i
H\left(\frac{\myreqbar_i}{\Rboubar_i}\right). \hspace{2.7cm}
\end{align*}
Note that 
\begin{equation*}
\lim_{\eps_2 \rightarrow 0}{\Gfunc'(\gamtwobar,
\widebar{y}_1,\ldots,\widebar{y}_{\delp},\eps_2)}=\lim_{\sbar_1
\rightarrow 0, \gambar \rightarrow 0}\Gfunc(\sbar_1,\gambar,\gamtwobar,
\widebar{y}_1,\ldots,\widebar{y}_{\delp},\eps_2),
\end{equation*}
where the limit is taken by setting $\gambar=\Theta (\sbar_1)$ (which will be true by concentration).
Therefore, we have
\begin{equation*}
\lim_{\eps_2 \rightarrow 0}{F'(\alpha,\eps_2)}\le F(\alpha).
\end{equation*}
Consequently, if $F(\alpha)<0$, it then follows that $\lim_{\eps_2
\rightarrow 0}{F'(\alpha,\eps_2)} < 0$. By continuity, there exists
some value $\eps_2>0$ such that $F'(\alpha,\eps_2) < 0$; for this
value of $\eps_2$, the probability $\Qprob[\BadEventB(A_3)\, \mid \,
\ReqTail(\epsone)]$ decreases exponentially in $n$.

\end{document}